\documentclass[english]{IEEEtran}
\usepackage[T1]{fontenc}
\usepackage[latin9]{inputenc}
\usepackage{pifont}
\usepackage{amsthm}
\usepackage{amsmath}
\usepackage{amssymb}
\usepackage{graphicx}
\usepackage{epstopdf}
\usepackage{epstopdf}
\usepackage{paralist}
\usepackage[T1]{fontenc}
\usepackage{epsfig}
\usepackage{cite}
\usepackage{dsfont}
\usepackage{bbm}
\usepackage[small]{caption}
\usepackage{subfigure}	
\usepackage{float}
\usepackage{graphicx}
\usepackage{framed}
\usepackage{wrapfig}

\makeatletter

\theoremstyle{plain} \newtheorem{thm}{\protect\theoremname}
\theoremstyle{plain} 
\theoremstyle{plain} 
\theoremstyle{plain} 
\theoremstyle{definition} \newtheorem{rem}{Remark}

\newcommand\oprocendsymbol{\hbox{$\square$}} 
\newcommand\oprocend{\relax\ifmmode\else\unskip\hfill\fi\oprocendsymbol}

\providecommand{\lemmaname}{Lemma}
\providecommand{\theoremname}{Theorem}
\providecommand{\examplename}{Example}
\providecommand{\corollaryname}{Corollary}

\@ifundefined{showcaptionsetup}{}{
 \PassOptionsToPackage{caption=false}{subfig}}
\usepackage{subfig}
\AtBeginDocument{
  
}

\makeatother

\usepackage{babel}

\makeatletter
\adddialect\l@ENGLISH\l@english
\makeatother
\begin{document}
\title{Uncovering Droop Control Laws Embedded Within the Nonlinear Dynamics of Van der Pol Oscillators}

\author{Mohit Sinha, Florian D\"{o}rfler, \emph{Member, IEEE}, \linebreak Brian B. Johnson, \emph{Member, IEEE},  and Sairaj V. Dhople, \emph{Member, IEEE}
\thanks{M. Sinha and S. V. Dhople are with the Department of Electrical and Computer Engineering at the University of Minnesota, Minneapolis, MN (email: \texttt{sinha052, sdhople@UMN.EDU}). F. D\"{o}rfler is with the Automatic Control Laboratory at ETH Z\"{u}rich, Z\"{u}rich, Switzerland (email: \texttt{dorfler@ETHZ.CH}). B. B. Johnson is with the Power Systems Engineering Center at the National Renewable Energy Laboratory (NREL), Golden, CO (email: \texttt{brian.johnson@NREL.GOV}) and his work was supported by the Laboratory Directed Research and Development Program at NREL.}}

\maketitle
\begin{abstract}
This paper examines the dynamics of power-electronic inverters in islanded microgrids that are controlled to emulate the dynamics of Van der Pol oscillators. The general strategy of controlling inverters to emulate the behavior of nonlinear oscillators presents a compelling time-domain alternative to ubiquitous droop control methods which presume the existence of a quasi-stationary sinusoidal steady state and operate on phasor quantities. We present two main results in this work. First, by leveraging the method of periodic averaging, we demonstrate that droop laws are intrinsically embedded within a slower time scale in the nonlinear dynamics of Van der Pol oscillators.  Second, we establish the global convergence of amplitude and phase dynamics in a resistive network interconnecting inverters controlled as Van der Pol oscillators. Furthermore, under a set of non-restrictive decoupling approximations, we derive sufficient conditions for local exponential stability of desirable equilibria of the linearized amplitude and phase dynamics. 
\end{abstract}

\section{Introduction}

\IEEEPARstart{A}{n} islanded inverter-based microgrid is a collection of heterogeneous DC energy resources, e.g., photovoltaic (PV) arrays, fuel cells, and energy-storage devices, interfaced to an AC electric distribution network and operated independently from the bulk power system. Energy conversion is typically managed by semiconductor-based power-electronic voltage-source inverters. The goal of decentralized real-time control is to regulate the inverters' terminal-voltage amplitude and frequency to realize a stable power system while achieving a fair and economic sharing of the network load.

\begin{figure}[t!]
\centering \includegraphics[width = 3.25in]{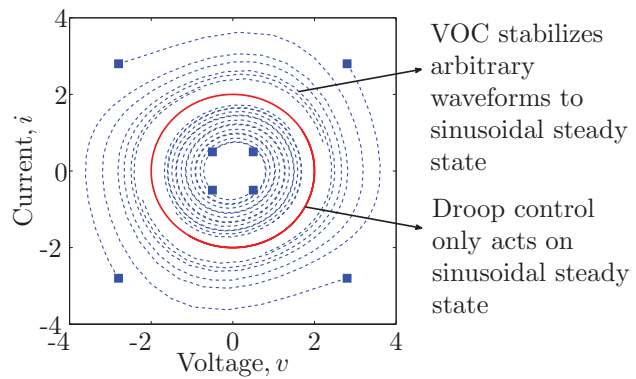}
\caption{VOC stabilizes arbitrary initial conditions to a sinusoidal steady state, while droop control acts on phasor quantities; only well defined in the sinusoidal steady state. One contribution of this work is to determine a set of parametric correspondences such that both approaches admit identical dynamics in sinusoidal steady state.}
\label{fig:VOCversusDroop} \vspace{-0.25in}
\end{figure}

The vast majority of academic and industrial efforts approaches the real-time control challenge by means of \emph{droop control}~\cite{Chandorkar-1993,Zhong13,Pogaku_2007,Bidram_2012}. Drawing from the control of synchronous generators in bulk power systems, droop control linearly trades off the active and reactive power injection with the inverters' terminal-voltage amplitude and frequency. In this paper, we focus on a communication-free decentralized control strategy wherein islanded inverters are regulated to mimic the dynamics of nonlinear limit-cycle oscillators~\cite{SD-SynchTCAS1-2014,SD-SynchJPV-2014,Dhople-Allerton-2013,SD-SynchTPELS-2014}. This method is inspired by synchronization phenomena in complex networks of coupled oscillators, and is termed \emph{Virtual Oscillator Control} (VOC). In general, VOC is executed by programming nonlinear differential equations of limit-cycle oscillators onto inverters' microcontrollers, and utilizing pertinent sinusoidally varying oscillator dynamic states to construct the pulse-width modulation (PWM) control signal. It is worth emphasizing that VOC constitutes a \emph{time-domain} approach and stabilizes arbitrary initial conditions to a sinusoidal steady state. As such, it is markedly different from droop control which operates on phasor quantities and presumes the existence of a quasi-stationary AC steady state; see Fig.~\ref{fig:VOCversusDroop}. See also~\cite{TorresHespanhaMoehlisSep13,TorresHespanhaMoehlisJul12} for similar time-domain control strategies.

Extending our previous efforts in~\cite{SD-SynchTCAS1-2014,SD-SynchJPV-2014,Dhople-Allerton-2013,SD-SynchTPELS-2014} where we focused on deadzone oscillators, in this paper we investigate the voltage dynamics of power-electronic inverters controlled to emulate the dynamics of Van der Pol oscillators (essentially, smooth cubic polynomial realizations of deadzone oscillators). Unless stated otherwise, in subsequent discussions where we reference VOC, we imply the control strategy is implemented with Van der Pol oscillators; also, inverters controlled with this approach are termed virtual-oscillator controlled (VO-controlled) inverters. Coupled Van der Pol oscillators tend to synchronize without any external forcing~\cite{RHR-PJH:80,Strogatz_Book01}, and hence utilizing them as virtual oscillators for inverter control is an effective strategy for realizing a stable AC microgrid.

We provide two main contributions in this paper: First, a correspondence is established between VOC and droop control by obtaining conditions under which the respective voltage dynamics at the inverter terminals---close to the sinusoidal steady state---are identical. To bridge the temporal gap between droop control and VOC, we average the periodic nonlinear oscillator dynamics to focus on AC-cycle time scales~\cite{HKK:02}. In addition to yielding insightful circuit-theoretic interpretations for droop control, our analysis highlights the choice of design parameters that ensure VO-controlled inverters mimic the behavior of droop-controlled inverters close to the quasi-stationary sinusoidal steady state and vice versa (see Fig.~\ref{fig:VOCversusDroop}). This allows us to leverage insights on the optimal choice of droop coefficients~\cite{FD-JWSP-FB:14a} to design VO-controlled inverters that achieve load sharing or economic optimality in steady state.

The second contribution of this work is to demonstrate the convergence of the averaged terminal-voltage amplitude and phase dynamics of VO-controlled inverters in resistive networks using a gradient-sytem formulation in concert with LaSalle's invariance principle. Under a set of non-restrictive decoupling assumptions on the phase and amplitude dynamics---valid in unstressed networks with a nearly uniform voltage profile and approximately equal phase angles~\cite{kundur1994power,Chandorkar-1993, Chandorkar13, Zhong_Robust13, Guerrero_Hierarchy11, RM-AG-GL-FZ:09}---we also present sufficient conditions for  local exponential stability of potentially desirable equilibria of the linearized and averaged VO-controlled inverter dynamics.

Within the realm of analytical approaches that investigate stability and synchronization in this application domain, for the deadzone type oscillators and parallel-connected inverters considered in~\cite{SD-SynchTCAS1-2014,SD-SynchJPV-2014,Dhople-Allerton-2013,SD-SynchTPELS-2014}, we utilized small-gain type arguments to prove synchronization; these results were generalized in terms of oscillator type and network topology recently in~\cite{SD-BJ-FD-AH:13} by leveraging structural and spectral properties of a network reduction procedure called Kron reduction~\cite{Dorfler-13}. Related work in~\cite{TorresHespanhaMoehlisSep13,TorresHespanhaMoehlisJul12} employed similar arguments based on incremental passivity. From a dynamical systems perspective, we establish a connection between limit-cycle oscillators (VO-controlled inverters) and phase oscillators (droop-controlled inverters) by means of coordinate transformations and averaging. For Van der Pol oscillators, similar connections and synchronization analyses date back to \cite{RHR-PJH:80} and have recently been surveyed in the tutorial~\cite{Mauroy-2012}. Additionally, averaging methods have recently been applied to study synchronization in Li\'{e}nard-type oscillators~\cite{tuna2012synchronization}, which include Van der Pol oscillators as a particular case. It is also worth mentioning that similar averaging methods have been applied to extract small-signal state-space models for DC-DC power-electronic converters~\cite{Krein-1989,sanders1991generalized,kimball2008singular,lehman1996switching,caliskan1999multifrequency}. Finally, we emphasize that the averaging analysis adopted here applies to general planar Li\'{e}nard-type limit-cycle oscillators which include Van der Pol oscillators as a particular case~\cite{Dhople_ACC_2015}.

Related to this work, for droop-controlled inverters in radial lossless microgrids under the assumption of constant voltage amplitudes, analytic conditions for proportional power sharing and synchronization have recently been derived by applying results from the theory of coupled oscillators in~\cite{Simpson-Porco_Synchronization12,FD-JWSP-FB:14a}. Conditions for voltage stability for a lossless parallel microgrid with one common load have been derived in~\cite{JWSP-FD-FB:13}. A decentralized linear matrix inequality-based control design for guaranteeing network stability considering variable voltage amplitudes and phase angles for meshed networks while accounting for power sharing has been described in~\cite{schiffer2013synchronization}.

The remainder of this manuscript is organized as follows. Section~\ref{sec:Preliminaries} establishes notation and relevant mathematical preliminaries. In Section~\ref{sec: Fundamentals}, we introduce droop control and VOC, and derive parametric conditions under which inverter dynamics controlled with the two approaches are identical. Next, in Section~\ref{sec: Amplitude Phase Dynamics}, we establish
global convergence of solutions for VO-controlled inverters in resistive networks; we also derive conditions for the exponential stability of linearized and decoupled amplitude and phase dynamics. Finally, we provide numerical simulations in Section~\ref{sec:Simulations}, and conclude the paper in Section~\ref{sec:Conclusions} by highlighting directions for future work.

\section{Notation and Preliminaries}
\label{sec:Preliminaries}

\subsection{Electrical System Fundamentals}
The nominal system frequency is denoted by $\omega$, and for the $j$th inverter, the instantaneous phase angle, $\phi_j$, evolves as
\begin{equation} \label{eq:phidef}
\frac{d\phi_j}{dt} = \omega + \frac{d\theta_j}{dt},
\end{equation}
where $\theta_j$ represents the phase offset with respect to the rotating reference frame established by $\omega$. Denote the instantaneous current injected by the $j$th inverter by $i_j(t)$ and its instantaneous terminal voltage by $v_j(t)$. Since we are primarily interested in harmonic signals, we parameterize the instantaneous voltage as $v_{j}(t) := r_j(t) \cos(\omega t + \theta_j(t))$, where $r_j(t)$ is the instantaneous terminal-voltage amplitude. We define the instantaneous active- and reactive-power injections \cite{Wang-Duarte,Peng-Lai}
\begin{align}
 \label{eq:PQinst}
P_{j}(t)&:= v_{j}(t) i_{j}(t) = r_{j}(t)\cos(\omega t+\theta_{j}(t)) i_j(t),\\
Q_{j}(t)&:= v_{j}\left(t - \frac{\pi}{2}\right) i_{j}(t) = r_{j}(t)\sin(\omega t+\theta_{j}(t)) i_j(t). \nonumber
\end{align}

Assuming the fundamental frequency of the current injected by the $j$th inverter is $\omega$, the average active and reactive power over an AC cycle (of period $2\pi/\omega$) are then given by
\begin{equation}
 \label{eq:PQaverage}
\overline{P}_j =\frac{\omega}{2\pi} \int_{s=0}^{\frac{2\pi}{\omega}} P_j(s) ds, \, \,
\overline{Q}_j =\frac{\omega}{2\pi} \int_{s=0}^{\frac{2\pi}{\omega}} Q_j(s) ds.
\end{equation}
In general, the time average of a periodic signal $u_j$ with period $T$ is denoted by $\overline{u}_j$, and defined as:
\begin{equation}
\overline{u}_j:=\frac{1}{T}\int_0^Tu_{j}(t) dt.
\end{equation}
Subsequent developments will leverage signals represented in the scaled time coordinates $\tau = \omega t$, and for the continuous-time signal $x$, we will denote $\dot{x} = \frac{d}{d \tau} x$.

\subsection{Mathematical Notation}
For the $N$-tuple, $\{x_1, \dots, x_N\}$, denote $x= [x_1, \dots, x_N]^\mathrm{T}$ to be the corresponding column vector; $(\cdot)^\mathrm{T}$ denotes transposition. The cardinality of the set $\mathcal{X}$ is denoted by $|\mathcal{X}|$; $[X]_{ij}$ isolates the entry in the $i$th row and $j$th column of matrix $X$. $\mathbb{R}^N$ is the space of $N\times1$ real-valued vectors, $\mathbb T^{N}$ is the $N$-torus. Given a scalar function $f(x)$, $\nabla_x f(x)$ returns the gradient $[\frac{\partial f}{\partial x_1}, \ldots, \frac{\partial f}{\partial x_n}]^\mathrm{T}$. Finally, $\mathrm{diag}\{x_1,\dots,x_N\}$ denotes a diagonal matrix with diagonal entries given by $x_1, \dots, x_N$. 

\section{Correspondence between Droop Control and VOC for Inverter Control}
\label{sec: Fundamentals}
In this section, we derive the droop coefficients under which the dynamics of droop control match VOC. We begin with a brief overview of droop control and VOC.

\subsection{Droop control}
For resistive networks, droop control linearly trades off frequency deviation versus reactive-power; and inverter terminal-voltage amplitude versus active-power~\cite{Zhong13,Zhong}:
\begin{equation}
\frac{d}{dt}\overline{\theta}_{j}= n_{j}\left(\overline Q_j - {\overline Q^{*}_{j}}\right),  \quad
\overline{r}_{j}- \overline{r}_{j}^* = m_{j}\left( \overline{P}_{j}^{*} - \overline{P}_{j}\right),
\label{eq: Droop Control}
\end{equation}
where ${\overline{Q}_{j}^{*}}$ and $\overline{P}_{j}^*$  are the per-phase average reactive-power and active-power setpoints, respectively; $\overline{r}_{j}^*$ is the terminal-voltage-amplitude setpoint; and $n_j, m_j \in \mathbb{R}_{>0}$ are reactive-power and active-power droop coefficients, respectively. As expressed in~\eqref{eq: Droop Control}, we assume that the droop laws are executed with AC-cycle averages of active and reactive power. To preserve the generality of the ensuing discussions, we disregard the dynamics of additional low-pass filters, voltage controllers, and current controllers in experimental implementations~\cite{Pogaku_2007}; however, these could be included in the analysis readily.

\begin{figure}[t!]
        \centering{
        \includegraphics[]{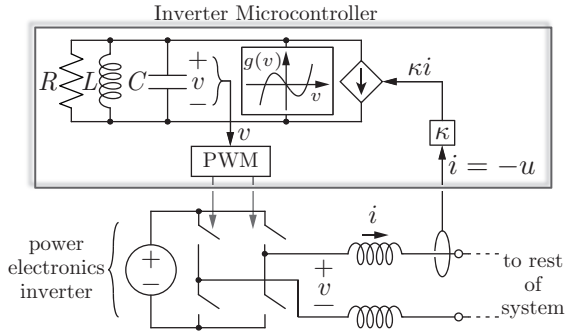} \label{Fig: Amp Error}
        \caption{Implementation of VOC for a single-phase power-electronic inverter. The Van der Pol oscillator is composed of a parallel $RLC$ circuit, and a nonlinear voltage-dependent current source, $g(v)$. The capacitor voltage is utilized as the PWM modulation signal.}
        \label{fig:ControlImplementation}}
\end{figure}

\subsection{VOC implemented with a Van der Pol Oscillator}
\label{sec:Oscillator Dynamics}
Consider the Van der Pol oscillator to constitute the virtual oscillator circuit for inverter control as shown in Fig.~\ref{fig:ControlImplementation}. The circuit implementation is composed of a parallel $RLC$ circuit and a nonlinear voltage-dependent current source, $g(\cdot)$. In the scaled time coordinates $\tau=t/\sqrt{LC}$, the dynamics of the oscillator are captured by the following:\footnote{For notational simplicity, we drop the subscript from electrical quantities and parameters that indexes the inverter in this section.}
\begin{equation}
	\ddot v -  \sqrt{\frac{L}{C}} \left(\sigma-\frac{1}{R}\right) \left( 1- \frac{3k}{(\sigma-\frac{1}{R})} v^{2} \right) \dot v + v = \kappa \sqrt{\frac{L}{C}} \dot u(\tau),
	\label{eq: VOC}
\end{equation}
where $u(\tau)$ is the current input to the Van der Pol oscillator (see Fig.~\ref{fig:ControlImplementation}), and $\kappa$ is the \emph{current gain}. In particular, the inverter output current is scaled by $\kappa$, and this is extracted from the Van der Pol oscillator that forms the inverter controller. The system in~\eqref{eq: VOC} can be compactly written as
\begin{equation}
	{
	\ddot v -  \varepsilon \alpha \bigl( 1- \beta v^{2} \bigr) \dot v + v = \kappa \varepsilon \dot u(\tau)
	}\,,
	\label{eq: VOC -- parameterized}
\end{equation}
by defining the following parameters:
\begin{equation}
\varepsilon :=\sqrt{\frac{L}{C}}, \quad \alpha := \sigma-\frac{1}{R}, \quad \beta := \frac{3k}{(\sigma-\frac{1}{R})}.
\label{eq: parameters}
\end{equation}
With this notation in place, the nonlinear voltage-dependent current source is a cubic polynomial, $g(v) = v - \beta (v^3 /3)$ (see Fig.~\ref{fig:ControlImplementation}). Li\'{e}nard's condition \cite{Strogatz_Book01} for ensuring a stable limit cycle in the system \eqref{eq: VOC -- parameterized} requires positive damping at the origin, i.e., $\alpha = \sigma-{1}/{R} > 0$. In the so-called quasi-harmonic limit, i.e., $\varepsilon \searrow 0$, the model~\eqref{eq: VOC -- parameterized} reduces to a forced harmonic oscillator with unit frequency. In the original time scale $t= \tau \sqrt{LC}$, this natural frequency of oscillation is $1/\sqrt{LC}$. By standard regular perturbation arguments \cite[Theorem 10.1]{HKK:02}, this correspondence can also be made for $\varepsilon$ sufficiently small.  In subsequent developments, with reference to~\eqref{eq:phidef}, and to compare the droop-control system~\eqref{eq: Droop Control} and the VOC system~\eqref{eq: VOC}, we set $\omega = 1/\sqrt{LC}$.

We begin by establishing a state-space model in Cartesian coordinates; choosing a scaled version of the inductor current and  capacitor voltage as states, $x:= \varepsilon i_L$, and $y := v$, we get
\begin{equation}
		\dot x = y, \quad \dot y = - x + \varepsilon \alpha g(y) + \varepsilon \kappa u(\tau).
\label{eq: VdP}
\end{equation}
Next, we transform the model~\eqref{eq: VdP} to polar coordinates by defining $x=r \sin(\phi)$ and $y=r \cos(\phi)$. We recover the following dynamics in polar coordinates:\footnote{This bijective change of coordinates is well defined (and leads to smooth dynamics) whenever $r \neq 0$ or equivalently $[x,y]^\mathrm{T} \neq 0$. In Theorem~\ref{Theorem: Global convergence of VO-controlled dynamics}, we establish well-posedness conditions focused on convergence of the amplitude dynamics to an equilibrium that excludes the origin.}
\begin{align}
\begin{split}
	\dot r &= \varepsilon \left( \alpha g\bigl(r\cos(\phi)\bigr) + \kappa u(\tau) \right) \cos(\phi),  \\
	\dot \phi &= 1 - \varepsilon \left(\frac {\alpha}{r} g\bigl(r\cos(\phi)\bigr) +  \frac{\kappa u(\tau)}{r} \right) \sin(\phi).
	\end{split}
\label{eq: VdP -- polar}%
		\end{align}
In ensuing discussions, we will leverage~\eqref{eq: VdP -- polar} written in the original time coordinates, with the nominal frequency of oscillation, $\omega = 1/\sqrt{LC}$, and phase offset as defined in~\eqref{eq:phidef}:
\begin{align}
\label{eq:polar_orig}
	\frac{dr}{dt} &= \frac{1}{C} \left( \alpha g\bigl(r\cos(\omega t + \theta)\bigr) + \kappa u(t) \right) \cos(\omega t + \theta),  \\
	\frac{d \theta}{dt} &=\omega -  \left(\frac{\alpha}{rC} g\bigl(r\cos(\omega t + \theta)\bigr) + \frac{\kappa u(t)}{rC} \right) \sin(\omega t + \theta). \nonumber
\end{align}
		
\begin{rem}[Controller implementation]
Essentially,~\eqref{eq:polar_orig},~\eqref{eq: Droop Control} describe the controller dynamics of the per-phase equivalent circuit at the inverter terminals; the signal $v = y = r \cos(\phi)$ can be utilized for control of single-phase inverters~\cite{SD-SynchTPELS-2014} (Fig.~\ref{fig:ControlImplementation}). For three-phase settings, a balanced set of PWM modulation signals, $m_\mathrm{a},m_\mathrm{b},m_\mathrm{c}$ are obtained as follows:
\begin{equation}
\begin{bmatrix}m_\mathrm a\\m_\mathrm b\\m_\mathrm c\end{bmatrix} =\Sigma^\mathrm{T}\begin{bmatrix}r \cos(\phi)\\ r \sin(\phi)\end{bmatrix}, \quad \Sigma :=\begin{bmatrix} 1 & -\frac{1}{2} & -\frac{1}{2}\\0 & \frac{\sqrt{3}}{2} & - \frac{\sqrt{3}}{2} \end{bmatrix}.
\end{equation}
The matrix $\Sigma$ implements a coordinate transformation from polar to $\mathrm{abc}$ coordinates~\cite{Iravani_Book10,SD-SynchJPV-2014}.\oprocend
\end{rem}

\subsection{Uncovering Droop Laws in Averaged VOC Dynamics}
Consider two microgrids, each with $N$ identical inverters, identical network configurations and loads. All inverters in one microgrid are controlled with VOC~\eqref{eq:polar_orig}, and the inverters in the other are controlled with droop control~\eqref{eq: Droop Control}. For the $j$th inverter, denote the difference in voltage amplitudes and phase offsets in the two inverter-control strategies by
\begin{equation}
	e_r(t) = \bar r_{j} - r_{j}(t), \quad 
	e_\theta(t) = \bar \theta_{j}(t) - \theta_{j}(t)
	\label{eq: signal differences}
\end{equation}
where $\bar r_{j}$ and $\bar\theta_{j}(t)$ are the amplitudes and phases as used in droop control \eqref{eq: Droop Control}, and  $r_{j}(t)$ and $\theta_{j}(t)$ those in VOC \eqref{eq:polar_orig}.

In the following, we analyze how the droop laws and coefficients should be designed so that the difference in the phase dynamics and steady-state equilibrium voltage profile of the two sets of inverters (controlled with VOC and droop) is of order $\mathcal{O}(\varepsilon)=\mathcal{O}(\sqrt{L/C})$. To bridge the time-scale separation between VOC (that is implemented in real-time) and droop control (that presumes the existence of a quasi-stationary sinusoidal steady state), we average the VOC dynamics~\eqref{eq:polar_orig} (a detailed derivation is provided in Step 1 of the proof to Theorem~\ref{thm:Correspondence} below) to arrive at the following description:
\begin{subequations} \label{eq:avgpolar}
\begin{align}%
	\frac{d}{dt}{\overline r}_j & = \frac{\alpha}{2C}\left(\overline r_j-\frac{\beta}{4}\overline r_j^3\right) -\frac{\kappa_j}{C \overline r_j} \overline P_j, \label{eq:avgpolar - amplitude} \\
	\frac{d}{dt}{\overline \theta}_j &= + \frac{\kappa_j}{C \overline r_j^2} \overline Q_j. \label{eq:avgpolar - phase}
\end{align}
\end{subequations}
The averaged VOC dynamics~\eqref{eq:avgpolar} enable us to compare the droop control laws in~\eqref{eq: Droop Control} with VOC~\eqref{eq:polar_orig}.

\begin{thm} [\bf Correspondence between Droop Control and VOC] \label{thm:Correspondence}
Consider two identical microgrids where all inverters in one microgrid are controlled with VOC~\eqref{eq:polar_orig}, and the inverters in the other are droop controlled~\eqref{eq: Droop Control}. Assume%
\begin{enumerate}

	\item[(A1)] unique solutions to the droop-controlled system \eqref{eq: Droop Control} and the averaged VOC system \eqref{eq:avgpolar} exist in a time interval $t \in [0,t^{*}]$ of strictly positive length.

	\item[(A2)] the average active power delivered by the $j$th inverter in sinusoidal steady state, $\overline P_{j,\mathrm{eq}}$, is bounded as
\begin{equation} \label{eq:PowerLimit}
0 < \kappa_j \overline P_{j,\mathrm{eq}} < \frac{\alpha}{2\beta}, 
\end{equation}
so that the average VOC dynamics \eqref{eq:avgpolar} admit a nonnegative amplitude equilibrium ${\overline{r}}_{j,\mathrm{eq}}$.
	\item[(A3)] both the VO-controlled microgrid~\eqref{eq:polar_orig} and the droop-controlled microgrid \eqref{eq: Droop Control} operate in steady state and the initial signal differences are of order $\varepsilon = \sqrt{L/C}$:
	\begin{equation*}
	e_r(0) \approx \mathcal{O}(\varepsilon)
	\quad\mbox{ and }\quad
	e_\theta(0) \approx \mathcal{O}(\varepsilon).
	\end{equation*}
\end{enumerate}
 Suppose the frequency-droop coefficient is picked as
\begin{equation} \label{eq:n}
n_{j}=\frac{\kappa_{j}}{\overline{r}_{j,\mathrm{eq}}^2 C} \,,
\end{equation}
and the average reactive-power setpoint is set to zero, ${\overline{Q}}_{j}^{*}=0$.
Suppose the voltage-droop coefficient is picked as
\begin{equation} \label{eq:m}
m_{j}=-\kappa_j \left(\alpha\left(\overline{r}_{j,\mathrm{eq}}-\frac{\beta}{2}\overline{r}^3_{j,\mathrm{eq}}\right)\right)^{-1} \,,
\end{equation}
and  the average active-power and amplitude setpoints are picked as ${\overline{P}}_{j}^{*}=\overline P_{j,\mathrm{eq}}$ and ${\overline{r}}_{j}^{*}={\overline{r}}_{j,\mathrm{eq}}$.
Then, there exists an $\varepsilon^{*}$, such that for all $0 < \varepsilon < \varepsilon^{*}$, for all $t \in [0,{t^*}]$
\begin{equation*}
	e_r(t) \approx \mathcal{O}(\varepsilon)
	\quad\mbox{ and }\quad
	e_\theta(t) \approx \mathcal{O}(\varepsilon).
\end{equation*}
\end{thm}
\noindent Assumption (A1) is guaranteed for $\overline r_{j}(0) >0$ due to local Lipschitz continuity; (A2) can be met by design; and (A3) is necessary for comparing the two strategies using averaging techniques.

The correspondences derived in Theorem \ref{thm:Correspondence}  are asymptotic results based on a perturbation and averaging analysis for sufficiently small $\varepsilon = \sqrt{L/C}$. However, a small $\varepsilon$ also implies a weak (nonlinear) viscous damping in \eqref{eq: VOC -- parameterized} and a slow convergence to the quasi-harmonic limit cycle. In Section~\ref{sec: convergence rate}, we show that the convergence rate is, in fact, inversely proportional to $\varepsilon$. Theorem~\ref{thm:Correspondence} and the above discussion indicate that the droop laws~\eqref{eq: Droop Control} are recovered from the VOC dynamics~\eqref{eq:polar_orig} only on \emph{slow} AC-cycle time scales, and when the dynamics of VO-controlled inverters are \emph{deliberately decelerated}. Hence, on the limit cycle, the decelerated VOC subsumes droop control, but it is much faster in general. Finally, the correspondences established in~\eqref{eq:n} and~\eqref{eq:m} are formally valid only on a bounded time horizon $[0,t^{*}]$. The findings can be extended to an unbounded time horizon provided that the averaged system is exponentially stable \cite{HKK:02}. In Section~\ref{sec: Amplitude Phase Dynamics}, we establish such exponential stability results.
\noindent \begin{IEEEproof}
The proof consists of three parts: an averaging analysis of VOC, a correspondence of the phase dynamics, and a correspondence of the steady-state voltage amplitudes.

\underline{\em 1) Averaging the VOC dynamics:}
We begin by averaging the dynamics~\eqref{eq: VOC -- parameterized}  of the VO-controlled microgrid. To this end, we first express~\eqref{eq:polar_orig} in the time coordinates $\tau = t / \sqrt{LC}$:\footnote{For notational simplicity, we drop the subscript $j$ from the variables $[r,\theta]^\mathrm{T}, [\overline r,\overline \theta]^\mathrm{T},\kappa, i, u$, indexing the $j$th inverter in equations~\eqref{eq:polar1}-\eqref{eq:averaging3}.}
\begin{align}
\label{eq:polar1}
\begin{split}
	\dot r &= \varepsilon \left( \alpha g\bigl(r\cos(\tau + \theta)\bigr) + \kappa u(\tau) \right) \cos(\tau + \theta),  \\
	\dot \theta &= - \varepsilon \left(\frac{\alpha}{r} g\bigl(r\cos(\tau + \theta)\bigr) + \frac{\kappa u(\tau)}{r} \right) \sin(\tau + \theta).
\end{split}
\end{align}
Note that the dynamical systems above are $2\pi$-periodic functions in $\tau$. In the quasi-harmonic limit $\varepsilon \searrow 0$, we can apply standard averaging arguments using $\varepsilon$ as the \emph{small parameter}, to obtain the averaged dynamics~\cite{HKK:02}:
\begin{align} 	\label{eq:averaging1}
\begin{bmatrix}\dot{\overline r} \\ \dot{\overline \theta}\end{bmatrix}
&= \frac{\varepsilon}{2\pi} \int_0^{2\pi} \alpha g\left(\overline r\cos(\tau + \overline \theta)\right) \begin{bmatrix}
\cos(\tau + \overline \theta)\\ -\frac{1}{\overline r}\sin(\tau + \overline \theta)
\end{bmatrix} d\tau \nonumber \\
&\quad + \frac{\varepsilon}{2\pi} \int_0^{2\pi} \kappa u(\tau)\begin{bmatrix}
\cos(\tau + \overline \theta)\\ -\frac{1}{\overline r}\sin(\tau + \overline \theta)
\end{bmatrix} d\tau  \\
&= \varepsilon \alpha \begin{bmatrix}
\frac{\overline{r}}{2} - \beta \frac{\overline{r}^3}{8} \\ 0
\end{bmatrix}  + \frac{\varepsilon}{2\pi} \int_0^{2\pi} \kappa u(\tau)\begin{bmatrix}
\cos(\tau + \overline \theta)\\ -\frac{1}{\overline r} \sin(\tau + \overline \theta)
\end{bmatrix} d\tau. \nonumber
\end{align}
The last line in~\eqref{eq:averaging1} follows from
\begin{align*}
&-\frac{\varepsilon}{2\pi \overline r} \int_0^{2\pi} \alpha g\left(\overline r\cos(\tau + \overline \theta)\right)\sin(\tau + \overline \theta)d\tau \\
&=\frac{\alpha \varepsilon}{2 \pi }\left( \left[\frac{1}{4} \cos(2\tau + 2\overline{\theta})\right]_0^{2\pi}+\frac{\beta \overline{r}^{2}}{3} \left[\cos^{4}(\tau + \overline{\theta})\right]_0^{2\pi} \right) = 0. \nonumber
\end{align*}
Transitioning~\eqref{eq:averaging1} from $\tau$ to $t$ coordinates, we get
\begin{equation*}
\label{eq:averaging2}
\begin{bmatrix}
\frac{d \overline{r}}{dt} \\ \frac{d \overline{\theta}}{dt}
\end{bmatrix} = \frac{\alpha}{C} \begin{bmatrix}
\frac{\overline{r}}{2} - \beta \frac{\overline{r}^3}{8} \\ 0
\end{bmatrix} +\frac{\kappa \omega}{2\pi C} \int_0^{\frac{2\pi}{\omega}} u(t)\begin{bmatrix}
 \cos(\omega t + \overline \theta)\\ -\frac{1}{\overline r} \sin(\omega t + \overline \theta)
\end{bmatrix} dt.
\end{equation*}
From Fig.~\ref{fig:ControlImplementation} we recognize that the current sourced by the Van-der-Pol oscillator is $i(t) = -u(t)$, and we get
\begin{align}
\begin{bmatrix}
\frac{d\overline{r}}{dt}\\ \frac{d\overline{\theta}}{dt}
\end{bmatrix}&= \frac{\alpha}{C} \begin{bmatrix}
\frac{\overline{r}}{2} - \frac{\beta \overline{r}^3}{8} \\ 0
\end{bmatrix} +\frac{\kappa \omega}{2\pi C} \int_0^{\frac{2\pi}{\omega}} \begin{bmatrix}
-i(t) \cos(\omega t + \overline \theta)\\ \frac{i(t)}{\overline r} \sin(\omega t + \overline \theta)
\end{bmatrix} dt \nonumber \\
&\hspace{-0.4in}= \frac{\alpha}{C} \begin{bmatrix}
\frac{\overline{r}}{2} - \frac{\beta \overline{r}^3}{8} \\ 0
\end{bmatrix} +\frac{\kappa \omega}{2\pi C} \int_0^{\frac{2\pi}{\omega}} \begin{bmatrix}
-\frac{i(t) \overline r}{\overline r}\cos(\omega t + \overline \theta)\\ \frac{i(t) \overline r}{\overline r^2} \sin(\omega t + \overline \theta)
\end{bmatrix} dt.
	\label{eq:averaging3}
\end{align}
Recalling the instantaneous and average active and reactive power definitions in~\eqref{eq:PQinst} and~\eqref{eq:PQaverage}, respectively, we observe that the averaged dynamics in~\eqref{eq:averaging3} are given by~\eqref{eq:avgpolar} (For details, see Appendix~\ref{sec:Avg}). 

Under assumptions (A1), (A2), and (A3), by standard averaging arguments \cite[Theorem 10.4]{HKK:02}, there exists an $\varepsilon_{1}^{*}$ sufficiently small so that
for all $0 < \varepsilon < \varepsilon_{1}^{*}$, the solution of the averaged VOC dynamics \eqref{eq:avgpolar} is $\mathcal O(\varepsilon)$ close to the solution of the original VOC dynamics \eqref{eq:polar_orig} for times $t \in [0,t^{*}/\varepsilon]$. We proceed by comparing the averaged VOC system \eqref{eq:avgpolar} with the droop control system \eqref{eq: Droop Control}.

\underline{\em 2) Correspondence of phase dynamics:}
We first study the phase dynamics \eqref{eq:avgpolar - phase}. The VOC system \eqref{eq:polar_orig} is assumed to evolve in quasi-stationary sinusoidal steady state with a small initial (at time $t=0$) $\mathcal{O}(\varepsilon)$ difference from the harmonic droop signals. Recall that in the quasi-harmonic limit, there exists an $\varepsilon_{2}^{*}$ sufficiently small so that for all $0 < \varepsilon < \varepsilon_{2}^{*}$, the solution of the VOC dynamics \eqref{eq:polar_orig} is $\mathcal O(\varepsilon)$ close to the solution of a harmonic oscillator with radius $\overline{r}_{j,\mathrm{eq}}$ for $t \in [0,t^{*}]$; see \cite{HKK:02,Mauroy-2012}. In particular, for $t \in [0,t^{*}]$, the solution ${\overline \theta}_j(t)$ of the averaged phase dynamics \eqref{eq:avgpolar - phase} is  $\mathcal O(\varepsilon)$ close to the solution of
\begin{equation*}
		\frac{d}{dt}{\overline \theta}_j = \frac{\kappa_j}{C \overline{r}_{j,\mathrm{eq}}^2} \overline Q_j
		\label{eq:avgpolar - phase - fixed radius}\,,
\end{equation*}
where we {disregard the amplitude dynamics~\eqref{eq:avgpolar - amplitude}, and} replace $\overline r_j(t)$ in \eqref{eq:avgpolar - phase} by $\overline{r}_{j,\mathrm{eq}}$ (whose closed form is discussed below).

For the following arguments, let $0 \leq \varepsilon \leq  \min \{ \varepsilon_{1}^{*},\varepsilon_{2}^{*}\}$. Observe that the phase dynamics of a droop-controlled inverter~\eqref{eq: Droop Control} correspond with the AC-cycle-averaged dynamics of a VO-controlled inverter~\eqref{eq:polar_orig}---up to an order $\mathcal O(\varepsilon)$ mismatch---if we pick the reactive-power setpoint, $\overline Q_j^*$, and the frequency-droop coefficient, $n_j$, as follows:
\begin{equation}
\overline{Q}_j^{*} = 0, \quad n_{j}=\frac{\kappa_{j}}{\overline{r}_{j,\mathrm{eq}}^2 C}.
\label{eq: reactive droop gain}
\end{equation}

\underline{\em 3) Correspondence of amplitude dynamics:}
Next, we consider the amplitude dynamics \eqref{eq:avgpolar - amplitude} and its equilibrium terminal-voltage profile. For the network of VO-controlled inverters, the steady-state voltage profile is recovered from the solution of the following $N$ nonlinear equations:
 \begin{equation}
	0 = \frac{\alpha}{2C}\bigg{(}  \overline{r}_{j,\mathrm{eq}} - \frac{\beta}{4} \overline{r}_{j,\mathrm{eq}}^{3}\bigg{)}  - \frac{\kappa_j \overline{P}_{j,\mathrm{eq}}}{C \overline{r}_{j,\mathrm{eq}}}, \quad \forall j = 1,\dots,N.
\label{eq:VOC Amplitude Steady State}
\end{equation}
Rearranging terms in~\eqref{eq:VOC Amplitude Steady State}, we get the following power-balance condition for the $j$th inverter
\begin{equation}
\frac{\alpha \beta}{8} \overline{r}_{j,\mathrm{eq}}^4 - \frac{\alpha}{2}{\overline{r}_{j,\mathrm{eq}}^2} + \kappa_j \overline{P}_{j,\mathrm{eq}} = 0.
\end{equation}
The positive roots of the above equation are given by
\begin{equation}
\overline r_{j,\mathrm{eq}} = \left[\frac{2\alpha \pm 2 \sqrt{\alpha^2 - 6 k \kappa_j \overline P_{j,\mathrm{eq}}}}{3k} \right]^{\frac{1}{2}},
\label{eq: roots}
\end{equation}
where we have used the fact that $\alpha \beta = 3k$ (see~\eqref{eq: parameters}). Notice that these two roots are real-valued if and only if~\eqref{eq:PowerLimit} holds. Around the high-voltage solution of~\eqref{eq: roots}, (denoted by $\overline{r}_{j,\mathrm{eq}}$ with a slight abuse of notation), the sensitivity of the active-power injection with respect to a change in amplitude is:
\begin{equation}
\kappa_j\frac{d\overline{P}_{j,\mathrm{eq}}}{d \overline{r}_{j,\mathrm{eq}}}=\alpha\left(\overline{r}_{j,\mathrm{eq}}-\frac{\beta}{2}\overline{r}^3_{j,\mathrm{eq}}\right), \quad \forall j = 1,\dots, N.
\label{eq:dPdr}
\end{equation}
In Theorem~3, we prove that this high-voltage solution is exponentially stable. Equation \eqref{eq:dPdr} can be placed in correspondence with the amplitude dynamics of a droop-controlled inverter~\eqref{eq: Droop Control}. By an analogous reasoning as for the phase dynamics, there exists an $\varepsilon_{3}^{*}$ sufficiently small so that for all $0 < \varepsilon < \varepsilon_{3}^{*}$, the solution ${\overline r}_j(t)$ of the averaged amplitude dynamics \eqref{eq:avgpolar - amplitude} satisfies---up to an $\mathcal O(\varepsilon)$ mismatch---the conditions of the stationary solution \eqref{eq:dPdr} (with fixed radius $\overline{r}_{j,\mathrm{eq}}$) for times $t \in [0,t^{*}]$.

For the following arguments, let $0 \leq \varepsilon \leq  \min \{ \varepsilon_{1}^{*},\varepsilon_{3}^{*}\}$. Observe that the amplitude dynamics of a droop-controlled inverter~\eqref{eq: Droop Control} correspond with that of a VO-controlled inverter in \eqref{eq:dPdr}---up to an order $\mathcal O(\varepsilon)$ mismatch---if we pick the active-power setpoint, $\overline P_j^*$, terminal-voltage setpoint, $\overline r_{j}^*$, and the voltage-droop coefficient, $m_j$, as follows:
\begin{equation*}
\overline{P}_j^{*} = \overline P_{j,\mathrm{eq}}, \,\, \overline r_{j}^* = \overline r_{j,\mathrm{eq}}, \,\,
m_{j} =-\kappa_j \left(\alpha\left(\overline{r}_{j,\mathrm{eq}}-\frac{\beta}{2}\overline{r}^3_{j,\mathrm{eq}}\right)\right)^{-1}.
\label{eq: active droop gain}
\end{equation*}
Finally, to complete the proof, let $\varepsilon^{*} = \min \{ \varepsilon_{1}^{*},\varepsilon_{2}^{*},\varepsilon_{3}^{*}\}$, and note that all arguments held for the time scales $[0,t^{*}/\varepsilon^*] \cap [0,t^{*}]$ which equals $[0,t^{*}]$ for $\varepsilon^{*}$ sufficiently small.
\end{IEEEproof}

\section{Stability of VOC Amplitude \& Phase Dynamics} \label{sec: Amplitude Phase Dynamics}
In this section, we investigate the stability of the averaged VOC voltage dynamics~\eqref{eq:avgpolar}. Our results are applicable to connected microgrid electrical networks with resistive interconnecting lines, and we place no restrictions on the network topology. Loads in the network are modeled as parallel connections of resistances and current sources/sinks (to simplify exposition, we refer to these as current sources subsequently).

\subsection{Microgrid Network Architecture} \label{sec: Microgrid}
We assume balanced three-phase operation and all electrical quantities referred henceforth are with respect to a per-phase equivalent network. The nodes of this per-phase equivalent electrical network are collected in the set $\mathcal{A}$, and branches (edges) are collected in the set $\mathcal{E} := \{(j,\ell)\} \subset \mathcal{A} \times \mathcal{A}$. Let $\mathcal{N}:=\{1,\dots,N\} \subseteq \mathcal{A}$ denote nodes that the inverters are connected to, and denote the set of \emph{internal nodes} as $\mathcal{I}:= \mathcal{A} \setminus \mathcal{N}$. Shunt loads---modeled as parallel combinations of resistances and/or constant (in a synchronous $\mathrm{dq}$-frame) current sources---are connected to $\mathcal{I}$.

Denote the vectors that collect the nodal current injections and node voltages in the network by $i_\mathcal{A}$ and $v_\mathcal{A}$, respectively. To be precise, $i_\mathcal{A}$ and $v_\mathcal{A}$ are real-valued functions of time. The coupling between the inverters is described by Kirchhoff's and Ohm's laws, which read in matrix-vector form as
\begin{equation}
i_\mathcal{A}=Q_\mathcal{A}v_\mathcal{A},\label{eq:i=Yv}
\end{equation}
where, entries of the \emph{conductance matrix} $Q_\mathcal{A} \in \mathbb{R}^{|\mathcal{A}|\times |\mathcal{A}|}$ are
\begin{equation}
[Q_\mathcal{A}]_{j\ell} := \left \{\begin{array}{ll}
{g}_{j} + \sum_{(j,k) \in \mathcal{E}} g_{jk} , & \mathrm{if} \,\,  j=\ell,\\
-g_{j\ell}, &\mathrm{if} \,\, (j,\ell)\in \mathcal{E},\\
0, & \mathrm{otherwise}, \end{array} \right.
\label{eq:Ymatrix}
\end{equation}
with $g_{j} \in \mathbb{R}_{\geq 0}$ denoting the shunt (load) conductance at node $j$, and $g_{j\ell} = g_{\ell j} \in \mathbb{R}_{\geq 0}$ the conductance of the line $(j,\ell)$.

Let $i \!=\! [i_1,\dots,i_N]^\mathrm{T}$ and $v \!=\! [v_1,\dots,v_N]^\mathrm{T}$ be the vectors of inverter current injections and terminal voltages, respectively, and let $i_\mathcal{I}$ and $v_\mathcal{I}$ be the vectors collecting the current injections and nodal voltages for the interior nodes.\footnote{We drop the subscript $\mathcal{N}$ when referring to the current and voltage vectors corresponding to the inverters to simplify notation.} Entries of $i_\mathcal{I}$ are non-zero only if the internal nodes are connected to current sources. With this notation, we can rewrite~\eqref{eq:i=Yv} as
\begin{equation}
\begin{bmatrix} i \\ i_\mathcal{I} \end{bmatrix} = \begin{bmatrix} Q_{\mathcal{N}\mathcal{N}} & Q_{\mathcal{N}\mathcal{I}} \\ Q_{\mathcal{N}\mathcal{I}}^\mathrm{T} & Q_{\mathcal{I}\mathcal{I}} \end{bmatrix} \begin{bmatrix} v \\ v_\mathcal{I} \end{bmatrix}.
\label{eq:OhmsLawMatrix}
\end{equation}
Assuming that the submatrix $Q_\mathcal{II}$ is nonsingular,\footnote{This holds true in general for $RLC$ networks, except for some pathological cases, see~\cite{SD-BJ-FD-AH:13}. For the resistive networks we consider in this work, $Q_\mathcal{II}$ is always nonsingular due to irreducible diagonal dominance \cite{Dorfler-13}.} the second set of equations in~\eqref{eq:OhmsLawMatrix} can be uniquely solved for the interior voltages as  $v_\mathcal{I} = Q_\mathcal{II}^{-1}(i_\mathcal{I} - Q_{\mathcal{N}\mathcal{I}}^\mathrm{T}v)$. Using this, we obtain:
\begin{align}
i &= Q v +Q_\mathcal{NI}Q^{-1}_\mathcal{II}i_\mathcal{I}, \label{eq:OhmsLaw}
\end{align}
where the matrix $Q = \left(Q_{\mathcal{NN}} - Q_{\mathcal{NI}} Q_{\mathcal{II}}^{-1} Q_{\mathcal{NI}}^\mathrm{T} \right)$ is referred to as the \emph{Kron-reduced conductance matrix}. This model reduction through a Schur complement of the conductance matrix is known as \emph{Kron reduction}~\cite{Dorfler-13}. With a slight abuse of notation, we denote the effective shunt-conductance load for the $j$th inverter by $g_j$ (note that this is given by the $j$th nonnegative row sum of the Kron-reduced conductance matrix $Q$), and the effective conductance of the $(j,\ell)$ line in the Kron-reduced electrical network by $g_{j\ell} = -[Q]_{j\ell}$ in all subsequent discussions. Additionally, the shunt current source at the $j$th inverter recovered after Kron reduction, given by the $j$th entry of the vector  $Q_\mathcal{NI}Q^{-1}_\mathcal{II}i_\mathcal{I}$, will be denoted by $\iota_j \cos(\omega t + \gamma_j)$, where $\iota_j$ is the amplitude of the current source, and $\gamma_j$ is the phase offset with respect to the rotating reference frame established by $\omega$. With this notation, the average real- and reactive-power injections for the $j$th inverter are given by~\cite{kundur1994power}:
\begin{align}
\overline P_{j}&=\frac{\overline r_{j}\iota_j}{2} \cos(\overline \theta_{j} - \gamma_j) + \frac{\overline r_j^2}{2} g_{jj} - \frac{ \overline r_{j}}{2}  \sum_{\ell=1, \ell \neq j}^{N}g_{j\ell} \overline r_{\ell}\cos( \overline \theta_{j\ell}),  \nonumber \\
     \overline Q_{j}&= \frac{\overline r_{j}\iota_j}{2} \sin(\overline \theta_{j} - \gamma_j) - \frac{\overline r_{j}}{2}  \sum_{\ell=1}^{N}g_{j\ell}\overline r_{\ell}\sin(\overline \theta_{j\ell}), \label{eq:avgPQ}
\end{align}
where we use the shorthand $\overline \theta_{j\ell}:= \overline \theta_j - \overline \theta_\ell$, and $\frac{\overline r_{j}\iota_j}{2} \cos(\overline \theta_{j} - \gamma_j)$ and $\frac{\overline r_{j}\iota_{j}}{2} \sin(\overline \theta_{j} - \gamma_j)$ are the active and reactive power drawn by the equivalent current source at the $j$th-inverter terminals (after Kron reduction). For these networks, we obtain the following well-posedness and global-convergence result.
\begin{thm}{(\bf Global convergence of VOC)}
\label{Theorem: Global convergence of VO-controlled dynamics}
Consider the interconnected averaged VOC dynamics~\eqref{eq:avgpolar} with real and reactive power injections given by~\eqref{eq:avgPQ}. Suppose that the terminal-voltage amplitudes are upper bounded by the open-circuit voltage, $\overline r^\mathrm{oc}:=\sqrt{4\alpha/3k}$.\footnote{The open-circuit voltage of the VO-controlled inverter is defined as the voltage obtained when no current is drawn from it. It is recovered from the high-voltage solution of~\eqref{eq: roots} by setting $\overline P_{j,\mathrm{eq}} = 0$.} Assume further, that the network and oscillator parameters satisfy $\forall j \in \mathcal{N}$
\begin{equation}
	\frac{16}{81}\left( {\alpha} - {\kappa_{j}g_{jj}} \right)^{3} \geq k \kappa_{j}^{2} \left( \iota_j+ \overline r^\mathrm{oc}\sum_{\ell=1,\ell \neq j}^{N}g_{j\ell} \right)^2  \,.
	\label{eq: key condition}
\end{equation}
Then, for all initial conditions $(\overline r_{0},\overline\theta_{0}) \in \mathbb{R}_{\geq 0}^{N} \times \mathbb{T}^N $ that satisfy
\begin{equation}
	 \overline r_j^\mathrm{low} := \sqrt{ \frac{4}{9k} \left({\alpha} - {\kappa_{j} g_{jj}} \right) } \leq \overline r_{0,j} \leq \overline r^\mathrm{oc}
	\,, \forall j \in \mathcal{N},
	\label{eq: chi condition}
\end{equation}
the dynamics \eqref{eq:avgpolar}, \eqref{eq:avgPQ} have positive radii $\overline r_{j}(t) \geq \overline r_j^\mathrm{low}$ for all $j \in \mathcal{N}$ and for all $t \geq 0$, and they ultimately converge to a set of equilibria as $t \to \infty$.
\end{thm}
We briefly discuss the assumptions in Theorem~\ref{Theorem: Global convergence of VO-controlled dynamics}. Condition \eqref{eq: key condition} assures that the radii $\overline r_{j}(t)$ remain greater than a strictly positive value $\overline r_j^\mathrm{low}$ given in \eqref{eq: chi condition}. Condition \eqref{eq: key condition}  is always guaranteed for sufficiently small current and resistive loads and a weakly coupled network, and it can be satisfied by choosing the ratio of design parameters $\alpha/\kappa_{j}$ sufficiently large. The proof of Theorem~\ref{Theorem: Global convergence of VO-controlled dynamics} relies on a gradient formulation of the system dynamics and LaSalle arguments:

\begin{IEEEproof}[Proof of Theorem~\ref{Theorem: Global convergence of VO-controlled dynamics}]
Inspired by~\cite{schiffer2013conditions}~\cite{ortega2002interconnection}, we begin by rewriting the system~\eqref{eq:avgpolar}, \eqref{eq:avgPQ} in gradient form as
\begin{subequations}%
\label{eq: PHS}
\begin{align}
	\dot{\overline r}_{j} &=: p_{j}(\overline r,\overline \theta) = - \nabla_{\overline r_j} H(\overline r, \overline \theta),
	\label{eq: PHS - amplitude} \\
	\dot{\overline \theta}_{j} &=: q_{j}(\overline r,\overline \theta) = - \frac{1}{\overline r_{j}^{2}} \nabla_{\overline \theta_j} H(\overline r, \overline \theta),
	\label{eq: PHS - phase}
\end{align}%
	\end{subequations}%
where $[\overline r,~\overline \theta]^\mathrm T = [\overline r_1,\dots,\overline r_N,\overline \theta_1,\dots,\overline \theta_N]^\mathrm T$, and the potential $H: \mathbb{R}^N_{\geq 0} \times \mathbb{T}^N \to \mathbb{R}$ is defined as
\begin{multline*}
H(\overline r, \overline \theta) := \sum_{j=1}^{N} \bigg{[} \frac{\alpha}{4C}\left(- \overline r_j^2 + \frac{\beta}{8} \overline r_j^4 \right)  + \frac{\kappa_j \iota_j}{2C}\overline r_j \cos(\overline \theta_j - \gamma_j)\\ +  \frac{\kappa_j}{4C} g_{jj}\overline r_j^2
 - \frac{\kappa_j}{2C} \sum_{\ell=1,\ell \neq j}^{N} \overline r_j \overline r_\ell g_{j\ell}\cos(\overline \theta_{j\ell}) \bigg{]}.
\end{multline*}
Notice that the phase dynamics  \eqref{eq: PHS - phase} are not defined for $\overline r_{j}=0$, and the notion of a {\em radius} is ill-posed whenever  $\overline r_{j}\leq 0$. Hence, we first establish conditions such that the radii remain greater than $\chi>0$, i.e., we seek conditions that ensure the set
\begin{equation*}
\Omega_{\chi}:= \left\{(\overline r,\overline \theta) \in \mathbb{R}_{\geq 0}^{N} \times \mathbb T^{N}:\, \chi \leq \overline r_{j} \leq \overline r^\mathrm{oc}  , \forall j \in \mathcal{N} \right\}
\end{equation*}
is positively invariant. To this end, we evaluate cases such that $p_{j}(\overline r,\overline \theta) \geq 0$ whenever $\left (\overline r, \overline \theta \right) \in \Xi_j \times \mathbb{T}^N$, where
\begin{equation}
 \Xi_j := \left\{ \overline r \in \mathbb{R}_{\geq 0}^N: \overline r_j = \chi_j,  \chi_\ell \leq \overline r_{\ell} \leq \overline r^\mathrm{oc}, \ell \neq j \right\},
\end{equation}
with $\chi_j$ and $\chi_\ell$ yet to be determined. In particular, $\forall j \in \mathcal N$
\begin{align*}
	 &p_{j}(\overline r,\overline \theta)|_{\left (\overline r, \overline \theta \right) \in \Xi_j \times \mathbb{T}^N}\\
	&=
	 \bigg{[}\frac{\alpha}{2C}\left(\overline r_j-\frac{\beta}{4} \overline r_j^3\right) -\frac{\kappa_j \iota_j}{2C} \cos( \overline \theta_{j} - \gamma_j) \\ &\quad-\frac{\kappa_j\overline r_j }{2C}g_{jj} +\frac{\kappa_j}{2 C}  \sum_{\ell=1,\ell \neq j}^{N}g_{j\ell} \overline r_{\ell}  \cos( \overline \theta_{j\ell}) \bigg{]}\bigg{|}_{\left (\overline r, \overline \theta \right) \in \Xi_j \times \mathbb{T}^N}
	\\&\geq
	 \frac{\alpha}{2C}\left(\chi_j-\frac{\beta}{4} \chi_j^3\right) - \frac{\kappa_j}{2C} \left(\iota_j + \chi_j g_{jj} + \overline r^\mathrm{oc} \sum_{\ell=1, \ell \neq j}^{N} g_{j\ell}\right) \\
	 &\geq 0,
\end{align*}
which holds if and only if there exists a $\chi_j \in \mathbb{R}_{>0}$ so that
\begin{align*}
	h_j(\chi_j) :=  \frac{\alpha\beta}{4} \chi_j^3 - \left( {\alpha} - {\kappa_{j} g_{jj}} \right) \chi_j + {\kappa_j \iota_j+\kappa_j\overline r^\mathrm{oc}\sum_{\ell=1,\ell \neq j}^N g_{j \ell}}
	\label{eq: parametric condition}
\end{align*}
is nonpositive. Since $h_j$ is a cubic polynomial with leading-order positive coefficient $\alpha \beta/4$, the question \emph{whether there is a $\chi_j>0$ so that $h_j(\chi_j)<0$} can be answered by calculating the positive maximum/minimum $\chi_j^{*}$ (the root of the equation $\partial h_j/\partial \chi_j = 0$) and verifying that $h_j(\chi_j^{*}) \leq 0$.\footnote{For $h(x) = a x^{3} - bx +c$, we obtain the extremal points by $0 = \frac{\partial h}{\partial x} = 3 ax^{2} - b$. If we assume that $a,b >0$, then the positive root is
	$x^{*} = \sqrt{\frac{b}{3 a}}$. We then obtain $h(x^{*}) =	a \frac{b}{3 a} \sqrt{\frac{b}{3 a}} - b \sqrt{\frac{b}{3 a}} + c.$ Notice that $h(x^{*}) \leq 0$ if and only if $ {a \frac{b}{3 a} - b} \leq - c \sqrt{\frac{3 a}{b}}$. This is equivalent to the condition $\frac{4}{27} b^{3} \geq ac^2$.}
The positive root $\chi_j^{*}$ is denoted by $\overline r_j^\mathrm{low}$ in \eqref{eq: chi condition} and $h_j(\overline r_j^\mathrm{low}) \leq 0$ if and only if \eqref{eq: key condition} holds true. Hence, under condition \eqref{eq: key condition}, we have positive invariance of the set
\begin{equation*}
\Omega:= \left\{(\overline r,\overline \theta) \in \mathbb{R}_{\geq 0}^{N} \times \mathbb T^{N}: \overline r^\mathrm{low}\leq \overline r_j^\mathrm{low} \leq \overline r_{j} \leq \overline r^\mathrm{oc}  , \forall j \in \mathcal{N} \right\},
\end{equation*}
where $\overline r^\mathrm{low}:= \min_{j\in \mathcal{N}} \overline r_j^\mathrm{low}$. Every trajectory originating in $\Omega$ remains in $\Omega$, i.e., $\overline r_{j}(t)$ is greater than $\overline r_j^\mathrm{low},\,\, \forall t\geq0$.

The level sets of $H(\overline r, \overline \theta)$ are closed (due to continuity), bounded in $\overline\theta$ (due to boundedness of the trigonometric nonlinearities), and radially unbounded in $\overline r$. Moreover, $H(\overline r, \overline \theta)$ is non-increasing along trajectories, since
\begin{align*}
	\dot H(\overline r, \overline \theta)
	&= - \sum_{j=1}^{N} \left(\nabla_{\overline r_j} H(\overline r, \overline \theta)) \right)^{2} +  \left(\frac{1}{\overline r_{j}} \nabla_{\overline \theta_j} H(\overline r, \overline \theta)) \right)^{2}
	\\
	&= - \sum_{j=1}^{N} p_{j}(\overline r,\overline \theta)^{2} +  \overline r_{j}^{2}q_{j}(\overline r,\overline \theta)^{2}
	\leq 0 \,.
\end{align*}
Thus, the sublevel sets of $H(\overline r, \overline \theta)$ are compact and forward invariant, and we conclude by LaSalle's invariance principle \cite[Theorem 4.4]{HKK:02} that the dynamics~\eqref{eq:avgpolar}, \eqref{eq:avgPQ} converge to the largest positively invariant set contained in
\begin{equation*}
\left\{(\overline r,\overline \theta) \in \Omega:\,  H(\overline r, \overline \theta) \leq H(\overline r_{0}, \overline \theta_{0}) \,,\, \dot H(\overline r, \overline \theta) = 0 \right\} \,,
\end{equation*}
where we incorporated the positive invariance of $\Omega$. The condition $\dot H(\overline r, \overline \theta) = 0$ identifies the set of equilibria and points of zero amplitude $\overline r_{j}=0$. Since the latter set is excluded from $\Omega$, all trajectories originating in $\Omega$ converge to the non-empty set of equilibria.
\end{IEEEproof}

Having demonstrated convergence and invariance of the averaged VO-controlled dynamics~\eqref{eq:avgpolar}, we next scrutinize the amplitude and phase dynamics under the standard \emph{decoupling} assumptions \cite{kundur1994power}. In particular, we assume the phase offsets (respectively, amplitudes) to be constant in the averaged amplitude (respectively, phase) dynamics in~\eqref{eq:avgpolar - amplitude} (respectively,~\eqref{eq:avgpolar - phase}). We are then able to derive sufficient conditions for the exponential stability of amplitude and phase dynamics.

\subsection{Amplitude Dynamics in Decoupled Settings}
Under the decoupling approximations described above, the phase offsets are fixed to their equilibrium values, i.e., $\overline \theta_j = \overline \theta_{j,\mathrm{eq}}, \forall j \in \mathcal{N}$; following which the terminal-voltage amplitude dynamics, recovered from~\eqref{eq:avgpolar - amplitude} and~\eqref{eq:avgPQ}, are given by:
\begin{align} \label{eq:decoupledr}
\dot{\overline r}_j &=\frac{\alpha}{2C}\left(\overline r_j-\frac{\beta}{4}\overline r_j^3\right) - \frac{\iota_j \kappa_j}{2C} \cos(\overline \theta_{j,\mathrm{eq}} - \gamma_j) \nonumber \\ &-\frac{\kappa_j}{2C} g_{jj}\overline r_{j}+ \frac{\kappa_j}{2C}  \sum_{\ell=1,\ell \neq j}^{N}g_{j\ell} \overline r_{\ell}\cos( \overline \theta_{j\ell,\mathrm{eq}}).
\end{align}

\begin{thm}[\bf Local exponential stability of decoupled amplitude dynamics]
\label{Theorem: Exponential Stability of amplitude dynamics}
Consider the decoupled terminal-voltage amplitude dynamics in~\eqref{eq:decoupledr}. Suppose each inverter is loaded according to~\eqref{eq:PowerLimit}.
If an equilibrium, $\overline{r}_{j,\mathrm{eq}}$, satisfies
\begin{equation} \label{eq:VoltageRegulation}
\overline r_j^\mathrm{low} <\overline{r}_{j,\mathrm{eq}} \leq \overline r^\mathrm{oc}, \,\, \forall j \in \mathcal{N},
\end{equation}
then it is locally exponentially stable.
\end{thm}

\begin{IEEEproof}[Proof of Theorem \ref{Theorem: Exponential Stability of amplitude dynamics}] For small perturbations about the equilibrium point $\overline r_\mathrm{eq} = [\overline r_{1,\mathrm{eq}},\dots, \overline r_{N,\mathrm{eq}}]^\mathrm{T}$ of~\eqref{eq:VOC Amplitude Steady State}, we express $\overline r = \overline r_\mathrm{eq} + \widetilde r$, where $\widetilde r := [\widetilde r_{1},\dots, \widetilde r_{N}]^\mathrm{T}$. Linearizing~\eqref{eq:decoupledr} around the equilibrium point (given by the solution of~\eqref{eq:VOC Amplitude Steady State}), $\overline r_\mathrm{eq}$, we get $\dot{\widetilde r} = K \Gamma \widetilde{r}$, 
where $K:=\mathrm{diag}\{\kappa_1,\dots,\kappa_N\}$. The diagonal entries of $\Gamma$ are
\begin{align*} 
[\Gamma]_{jj} &= \frac{\alpha}{2C\kappa_j}\left(1-\frac{3}{4}\beta \overline{r}_{j, \mathrm{eq}}^2\right) -\frac{1}{2C}\left(g_j + \sum_{\ell = 1, \ell \neq j}^{N} g_{j \ell} \right). 
\end{align*}
Furthermore, the matrix $\Gamma$ is irreducible (due to connectivity) and symmetric since
\begin{equation*} 
[\Gamma]_{j \ell} = [\Gamma]_{\ell j} = \frac{1}{2C}g_{j \ell}\cos(\overline \theta_{j,\mathrm{eq}} - \overline \theta_{\ell,\mathrm{eq}}).
\end{equation*}
If we ensure
\begin{equation} \label{eq:VoltageRegulationTemp}
\frac{\alpha}{2\kappa_j}\left(1-\frac{3}{4}\beta \overline{r}_{j, \mathrm{eq}}^2\right) -\frac{1}{2}g_{j}< 0,
\end{equation}
then $\Gamma$ is negative definite (due to strictly irreducible diagonal dominance \cite{RAH-CRJ:85}). By Sylvester's inertia theorem~\cite{Carlson_1963}, the inertia (i.e., the triple of positive, negative, and zero eigenvalues) of $\Gamma$ and $K\Gamma$ are identical since $\kappa_j > 0,\forall j \in \mathcal{N}$ and $K$ is positive definite. Consequently, $K\Gamma$ is negative definite, provided~\eqref{eq:VoltageRegulationTemp} is satisfied. The bounds in~\eqref{eq:VoltageRegulation} are obtained by rearranging terms in~\eqref{eq:VoltageRegulationTemp}. The upper bound in~\eqref{eq:VoltageRegulation} is the open-circuit voltage.\end{IEEEproof}

\subsection{Phase Dynamics in Decoupled Settings}
\label{sec: Phase Dynamics}
Under the decoupling assumptions, the terminal-voltage amplitudes are fixed to their equilibrium values, $\overline r_j = \overline r_{j,\mathrm{eq}}, \forall j \in \mathcal{N}$, and the phase dynamics \eqref{eq:avgpolar - phase} and~\eqref{eq:avgPQ} are given by:
\begin{equation}
\label{eq:decoupledtheta}
\dot{\overline \theta}_j = \frac{\kappa_j}{2 C \overline r_{j,\mathrm{eq}}} \left( \iota_j \sin(\overline \theta_{j} - \gamma_j) - \sum_{\ell=1,\ell \neq j}^{N}g_{j\ell}\overline r_{\ell,\mathrm{eq}}\sin(\overline \theta_{j\ell}) \right).
\end{equation}
Analysis of the decoupled phase dynamics~\eqref{eq:decoupledtheta} with coupled oscillator theory \cite{FD-MC-FB:12c,Dorfler-13-Synch} leads to the following result.

\begin{thm}[\bf Local exponential stability of decoupled phase dynamics]
\label{Theorem: Exponential stability of phase dynamics}
Consider the decoupled phase dynamics~\eqref{eq:decoupledtheta}. Assume that there exists an equilibrium $\overline{\theta}_{j,\mathrm{eq}}$ so that 
\begin{equation}
|\overline \theta_{j \ell,\mathrm{eq}}| < \pi/2
\quad\mbox{ and }\quad
|\overline \theta_{j,\mathrm{eq}}-\gamma_{j}| > \pi/2, \,\,\,\,\forall j,\ell \in\mathcal N.
\label{eq:stable theta}
\end{equation}
If there is at least one constant current load, then the equilibrium $\overline{\theta}_{j,\mathrm{eq}}$ is locally exponentially stable. Without constant current loads, the phase-synchronized equilibrium manifold $\overline{\theta}_{j,\mathrm{eq}} = \overline{\theta}_{\ell,\mathrm{eq}}$, for all $j,\ell \in \mathcal N$, is locally exponentially stable.
\end{thm}

Condition \eqref{eq:stable theta} identifies the equilibria corresponding to small reactive power flows (as suggested by the condition $|\overline \theta_{j \ell,\mathrm{eq}}| < \pi/2 $) and requires the local current sources to inject reactive power (as suggested by the condition $|\overline \theta_{j,\mathrm{eq}}-\gamma_{j}| > \pi/2$). Without current loads, the phase synchronization result perfectly matches our previous experimental results in \cite{SD-SynchTCAS1-2014,SD-SynchTPELS-2014}. 

\begin{IEEEproof}[Proof of Theorem \ref{Theorem: Exponential stability of phase dynamics}]
Linearization of~\eqref{eq:decoupledtheta} around the equilibrium point $\overline \theta_\mathrm{eq}$ yields $\dot{\widetilde \theta} = K \Theta M\widetilde{\theta}$, where $\overline \theta = \overline \theta_\mathrm{eq} + \widetilde \theta$, $K:=\mathrm{diag}\{\kappa_1/\overline r_{1,\mathrm{eq}},\dots,\kappa_N/\overline r_{N,\mathrm{eq}}\}$ and $M:=\mathrm{diag}\{\overline r_{1,\mathrm{eq}},\dots,\overline r_{N,\mathrm{eq}}\}$.
The matrix $\Theta$ is irreducible (due to connectivity), and symmetric with off-diagonal entries
\begin{equation*}  
[\Theta]_{j \ell} = [\Theta]_{\ell j} = \frac{ g_{j \ell} }{2C}\cos(\overline \theta_{j \ell,\mathrm{eq}}).
\end{equation*}
The diagonal entries of $\Theta$ are given by
\begin{equation*} 
[\Theta]_{jj}
:=
 \frac{\iota_{j}}{2 C \overline r_{j,\mathrm{eq}}}\cos(\overline \theta_{j,\mathrm{eq}}-\gamma_{j}) - \sum_{\ell=1 ,\ell \neq j}^{N}[\Theta]_{j \ell} .
\end{equation*}
Under assumption \eqref{eq:stable theta}, the off-diagonal entries $[\Theta]_{j \ell} $ are nonnegative, and all row sums are non-positive. If there is at least one constant current load, the associated row sum is strictly negative. Hence, $\Theta$ is irreducibly diagonal dominant (due to connectivity), and thus also nonsingular \cite[Corollary 6.2.27]{RAH-CRJ:85}. It follows that $\Theta$ is negative definite, and the equilibrium $\overline \theta_\mathrm{eq}$ is isolated and locally exponentially stable. In the absence of local current loads, the negative Jacobian, $-\Theta$, is a Laplacian matrix associated with an undirected and connected graph. For this matrix, the phase-synchronized equilibrium manifold is locally exponentially stable; see \cite[Theorem 5.1]{Dorfler-13-Synch} for details.

The eigenvalues of the matrix $(K\Theta)M$ are the same as $M(K\Theta)$ since $K, M$ are diagonal and $\Theta$ is symmetric. Again, by Sylvester's inertia theorem~\cite{Carlson_1963}, the inertia (i.e., the triple of positive, negative, and zero eigenvalues) of $\Theta$ and $MK\Theta$ are identical since $\kappa_j > 0, \overline r_j>0, \forall j \in \mathcal{N}$. Consequently, $K\Theta M$ is negative definite, and therefore, the phase dynamics are locally exponentially stable, provided that~\eqref{eq:stable theta} is satisfied.\end{IEEEproof}

\section{Reverse Engineering Droop Control, Convergence Rates, and Numerical Validation} \label{sec:Simulations}

Simulations in this section focus on corroborating the averaging analysis and the correspondence established with droop control. Additionally, we discuss the load-sharing capabilities afforded by VOC. Finally, we comment on implications of the quasi-harmonic limit, $\varepsilon \searrow 0$, on the VOC convergence speed.

\subsection{Correspondence between VOC and Droop Control}
\begin{figure}[t!]
\centering \includegraphics[scale = 0.28]{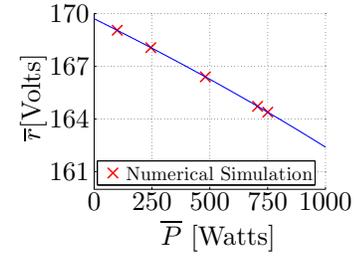}
\caption{Voltage-power characteristic~\eqref{eq: roots} for an inverter superimposed to time-domain simulations of the non-averaged nonlinear model~\eqref{eq:polar_orig} run out to steady state.}
\label{fig:VoltageRegulation}\vspace{-0.15in}
\end{figure}

First, we validate the averaging analysis by focusing on the expression in~\eqref{eq: roots}. In particular, the voltage-regulation curve for VOC (from~\eqref{eq: roots}) is plotted in Fig.~\ref{fig:VoltageRegulation} and the analytical expression is validated by comparison with simulations of the original nonlinear and non-averaged Van der Pol oscillator model~\eqref{eq:polar_orig} run out to steady state.

\begin{figure}[t!]
        \subfigure[]{\includegraphics[width=0.46\columnwidth]{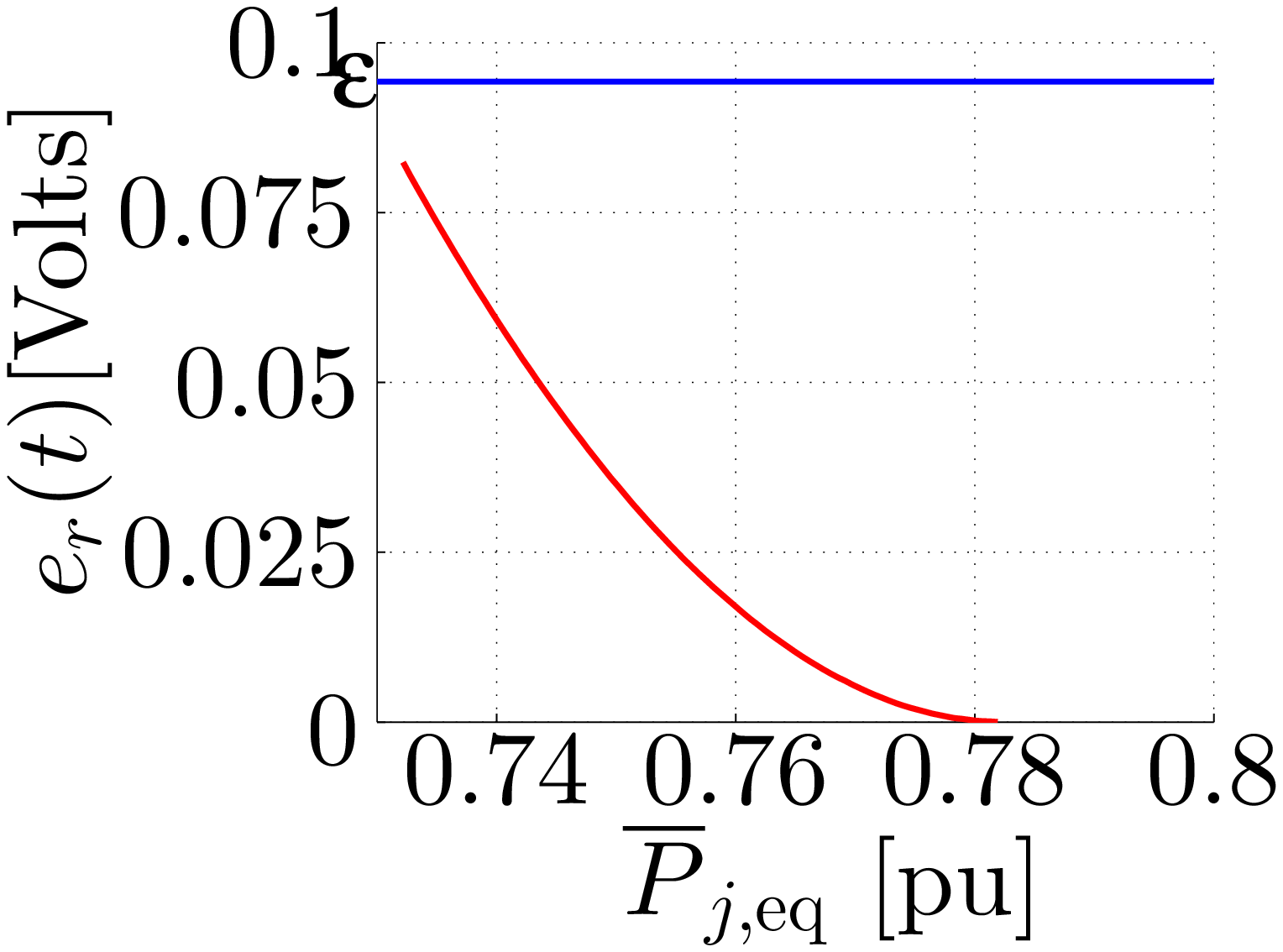}} \label{Fig: Amp Error}
        \subfigure[]{\includegraphics[width=0.49\columnwidth]{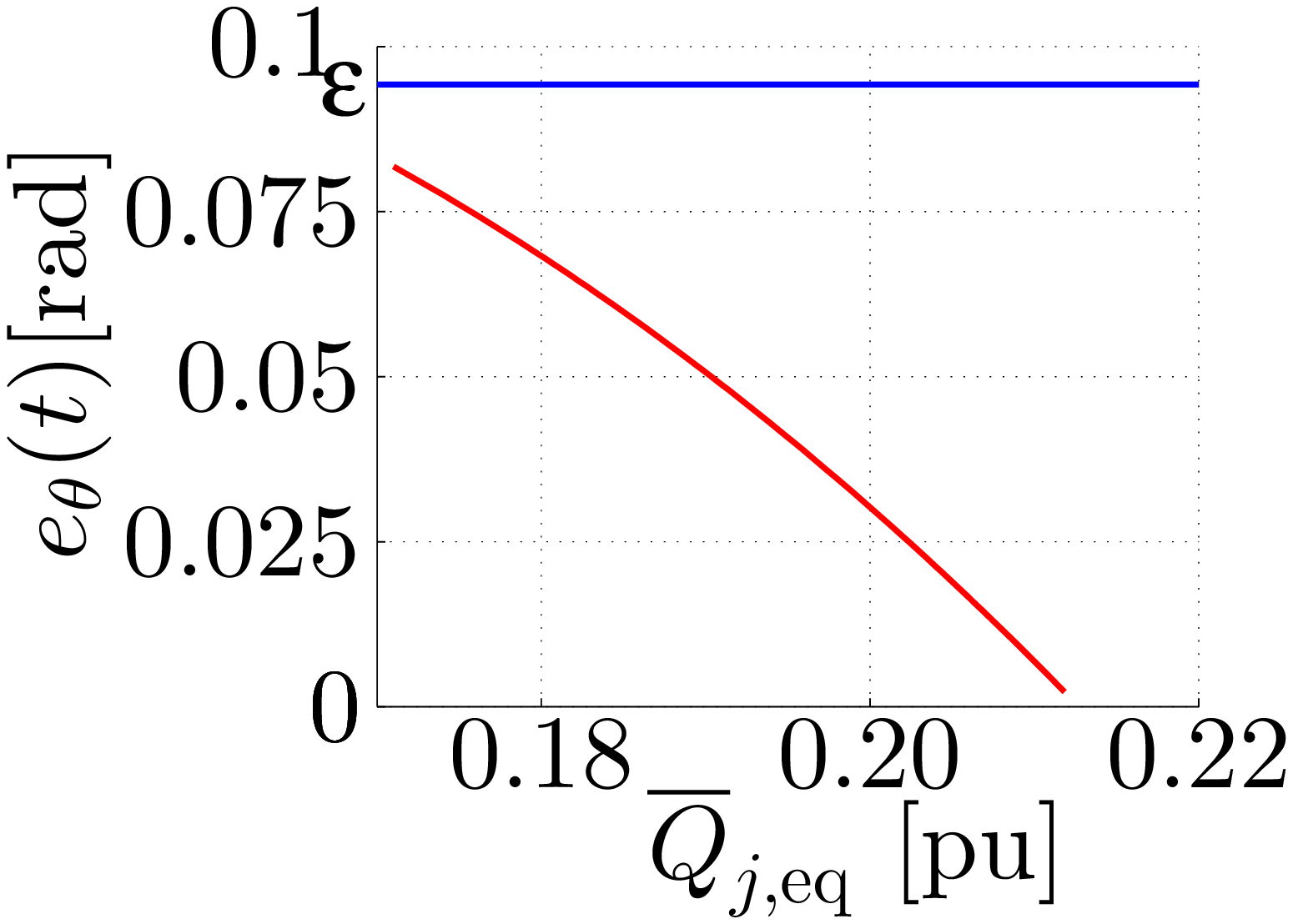}} \label{Fig: Freq Error}
        \caption{Differences in: (a) equilibrium-voltages and (b) phase-offsets when comparing VOC and droop control.}
        \label{Fig: Euclidean errors}\vspace{-0.2in}
\end{figure}

Next, we focus on the correspondences established between VOC and droop control. To this end, we model a single $15~\mathrm{kW}$ three-phase inverter connected to a load which draws a constant current at a lagging power factor of $0.85$. Suppose a Van der Pol oscillator-based controller (parameters are listed in the Appendix) is supplying $0.78~\mathrm{pu}$ active power and $0.21~\mathrm{pu}$ reactive power in steady state. A corresponding droop controller is derived using the expressions in~\eqref{eq:n} and~\eqref{eq:m}. Figure~\ref{Fig: Euclidean errors}(a) depicts $e_r(t)$ in steady state as the active power consumed by the load is varied. Figure~\ref{Fig: Euclidean errors}(b) depicts $e_\theta(t)$ recorded at time $t=2.5 \mathrm{s}$ as the reactive power consumed by the load is varied. Differences in both cases are of $\mathcal{O(\varepsilon)}$.

\subsection{Load Sharing and Economic Optimality}
Consider the microgrid setting where $N$ inverters are connected in parallel across a balanced three-phase load. In this case, droop control \eqref{eq: Droop Control} also achieves steady-state load sharing or economic optimality. For resistive networks, it is known \cite{Zhong13,Simpson-Porco_Synchronization12} that the steady-state reactive power injection $\overline{Q}_{j,\mathrm{eq}}$ from the $j$th inverter is proportional to its {rating $R_{j}$}, that is,
\begin{equation}
	\label{eq: load sharing}
	\frac{\overline{Q}_{j,\mathrm{eq}}}{{R_{j}}}
	=
	\frac{\overline{Q}_{\ell,\mathrm{eq}}}{{R_{\ell}}}
	\quad\forall\; j, \ell \in \{1,\dots, N\},
\end{equation}
provided the following hold:
\begin{equation*}
	\frac{\overline{Q}_{j}^{*}}{{R_{j}}}
	=
	\frac{\overline{Q}_{\ell}^{*}}{{R_{\ell}}}, \quad
	{n_{j}}{{R_{j}}}
	=
	{n_{\ell}}{{R_{\ell}}}
	\quad\forall\; j, \ell \in \{1,\dots, N\}.
\end{equation*}
Similarly, droop control can be designed to minimize an {unconstrained} economic dispatch of the reactive power injections
\begin{equation}
	\label{eq: econ dispatch}
	\displaystyle\min_{\{\overline{Q}_{j,\mathrm{eq}}\}_{j=1}^N} \quad \sum_{j=1}^{N} \lambda_{j} \overline{Q}_{j,\mathrm{eq}}^{2},
\end{equation}
with marginal costs $\lambda_{i} >0$ provided the reactive-power setpoints and droop coefficients are selected as follows~\cite{FD-JWSP-FB:14a}:\footnote{Note that the two objectives \eqref{eq: load sharing} and \eqref{eq: econ dispatch} and the associated droop gains coincide for $R_{j}/\lambda_{j} = R_{\ell}/\lambda_{\ell}$ for all $ j, \ell \in \{1,\dots, N\}$.}
\begin{equation} \label{eq:optimaln}
	\overline{Q}_{j}^{*}=0, \quad \frac{n_j}{\lambda_j}=\frac{n_\ell}{\lambda_\ell},
	\quad\forall\; j, \ell \in \{1,\dots, N\}.
\end{equation}
The correspondences established in Theorem~\ref{thm:Correspondence} allow us to translate these insights to the design of optimal current gains (i.e., the $\kappa$'s) in VO-controlled inverters \eqref{eq: VOC} to achieve optimality in terms of reactive-power production. In particular, leveraging~\eqref{eq: reactive droop gain}, and based on~\eqref{eq:optimaln}, the following design achieves an optimal dispatch of reactive power generation:
\begin{equation}
\frac{\kappa_{j}}{\overline{r}_{j,\mathrm{eq}}^2 \lambda_j} = \frac{\kappa_{\ell}}{\overline{r}_{\ell,\mathrm{eq}}^2 \lambda_\ell}, \quad j, \ell \in \{1,\dots,N\}.
\end{equation}
Similar load-sharing conditions have been obtained for inverters controlled as deadzone oscillators where all voltage waveforms perfectly synchronize (amplitude, frequency, and phase)~\cite{SD-SynchTPELS-2014}. In particular, picking the current gains $\kappa_{j}$ as
\begin{equation} \label{eq:sharing}
R_j \kappa_j = R_\ell \kappa_\ell 	,
\quad\forall\; j, \ell \in \{1,\dots, N\}\,,
\end{equation}
ensures that the current injections are shared proportionally \cite{SD-SynchTPELS-2014}, and thus---due to perfect synchronization of the voltage waveforms---the apparent power injections $\overline S_{j,\mathrm{eq}} = \overline P_{j,\mathrm{eq}} + \mathrm{j} \overline Q_{j,\mathrm{eq}} $ are shared proportionally in steady state:
\begin{equation}
\frac{\overline S_{j,\mathrm{eq}}}{R_{j}} = \frac{\overline S_{\ell,\mathrm{eq}}}{R_{\ell}},
\quad\forall\; j, \ell \in \{1,\dots, N\}.
\label{eq: inst power sharing}
\end{equation}
As a consequence, the average active and reactive injections are shared, and \eqref{eq: load sharing} is recovered as a special case. Results from Theorem~\ref{thm:Correspondence} allow us to extend load-sharing results for VO-controlled inverters from a setting with perfectly synchronized waveforms to more general frequency-synchronized waveforms. {Consider the closed-form high-voltage solution for the terminal-voltage amplitude of the $j$th inverter in~\eqref{eq: roots}. When the oscillators are identical, the terminal-voltage amplitudes synchronize if we pick the current gains as follows:\begin{equation}
\kappa_j \overline P_{j,\mathrm{eq}} = \kappa_\ell \overline P_{\ell,\mathrm{eq}}, \quad \forall j, \ell \in \{1,\dots, N\}.
\end{equation}

We simulate a case of power sharing between three identical VO-controlled inverters connected in a parallel configuration with current gains $\kappa=[2 \quad  2 \quad 1]^\mathrm{T}$. As shown in Fig.~\ref{fig:Power_Sharing}, two of the inverters share $25\%$ of the load while the third inverter provides $50\%$ of the load. A load step is applied at $t=1\mathrm{s}$ by doubling the active-power demand. The inverters support the load in the ratio of their ratings even after the load step.
\begin{figure}[t!]
\centering{
\includegraphics[width=0.46\columnwidth]{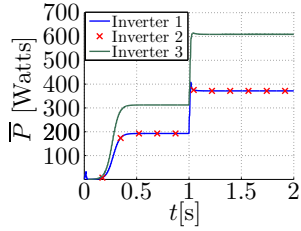}
}
\caption{Power sharing for 3 parallel VO-controlled inverters.}
\label{fig:Power_Sharing} \vspace{-0.2in}
\end{figure}

\subsection{Convergence Rate of a Van der Pol Oscillator}\label{sec: convergence rate}
In this section, we discuss the implication of the quasi-harmonic limit $\varepsilon \searrow 0$ on the time taken to converge to the limit cycle in an open-circuited Van der Pol oscillator, i.e., when setting the driving term $u = 0$. From~\eqref{eq: VdP -- polar}, we obtain
\begin{equation*}
\frac{dr}{d\phi}
=
\frac{ \varepsilon \alpha g\bigl(r\cos(\phi)\bigr) \cos(\phi) } {  1 - \varepsilon \frac {\alpha}{r} g\bigl(r\cos(\phi)\bigr)  \sin(\phi) }\,.
\end{equation*}
In the quasi-harmonic limit $\varepsilon \ll 1$, we apply the series expansion $\varepsilon/(1-\varepsilon \cdot c) = \varepsilon + \mathcal O(\varepsilon^{2})$ above to get
\begin{equation*}
\frac{dr}{d\phi}
=
\varepsilon \left( \alpha g\bigl(r\cos(\phi)\bigr) \right) \cos(\phi) + \mathcal O(\varepsilon^{2}).
\end{equation*}
Averaging the above dynamics yields (up to $\mathcal{O}(\varepsilon^2)$ terms):
\begin{equation} \label{eq: VdP - dr/dphi}
\frac{d \overline r}{d\overline \phi}=\frac{\alpha \varepsilon}{2}\bigg{(}  \overline r  - \frac{\beta}{4} \overline r^{3}\bigg{)}.
\end{equation}

Note that the locally stable equilibrium of the dynamics \eqref{eq: VdP - dr/dphi} is given by the open-circuit voltage, $\overline r_\mathrm{eq}=\overline r^\mathrm{oc}$. We integrate both sides of~\eqref{eq: VdP - dr/dphi}, arbitrarily setting the limits from $0.1 \overline r_\mathrm{eq}$ to $0.9 \overline r_\mathrm{eq}$ (without loss of generality). The arc length traced during this transition, $\phi_\mathrm{s}$, is given by the solution of:
\begin{equation*}
{\left[-\frac{1}{4}\log \overline r+\frac{1}{8}\log\lvert {4-\beta(\overline r})^2\rvert\right]}^{0.9 \overline r_\mathrm{eq}}_{0.1 \overline r_\mathrm{eq}}=-\frac{1}{8}\varepsilon\phi_\mathrm{s}.
\end{equation*}
Evaluating the limits of this integral, we recover $\phi_\mathrm{s}\approx 6\left({\varepsilon\alpha}\right)^{-1}$, which clearly indicates that the arc length $\phi_\mathrm{s}$ (proportional to a notion of convergence time to $\mathcal O(\varepsilon)$) traced before converging to the limit cycle is inversely proportional to $\varepsilon$.
\begin{figure}[t!]
\centering{
\includegraphics[width=0.46\columnwidth]{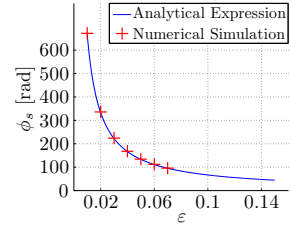}
}
\caption{Convergence rate of a Van der Pol oscillator.}
\label{Fig: Settling time}\end{figure}
Figure~\ref{Fig: Settling time} plots $\phi_\mathrm{s}$ as a function of $\varepsilon$. Results from simulations of the original unforced nonlinear dynamics \eqref{eq: VdP -- polar} (with $u=0$) are superimposed to demonstrate validity of the above analysis.

\section{Concluding Remarks} \label{sec:Conclusions}
For a system of power-electronic inverters controlled as Van der Pol oscillators, we characterized the voltage dynamics in polar coordinates to establish two key results: i)~we derived a set of parameters for which the dynamics of the Van der Pol oscillators match the classical droop laws close to sinusoidal steady state, and ii)~we established convergence of the Van der Pol oscillator dynamics to a set of potentially desirable equilibria. With this analysis, we are able to reverse-engineer droop control and ensure that VOC is compatible with secondary and tertiary control strategies developed for droop control. Extending the analysis to inductive networks while incorporating other load models and leveraging the averaged dynamics to design control strategies for general microgrid networks remain the focus of ongoing investigations.

\section*{Acknowledgments}
F. D\"{o}rfler would like to thank Rodolphe Sepulchre and Pierre Sacr\'{e}; and B. Johnson would like to thank Nathan Ainsworth for insightful discussions. 

\begin{appendix}
\subsection{Simulation Parameters}
\emph{Oscillator parameters:} $R = 10\,\Omega$, $L = 250\,\mathrm{\mu H}$, $C= 28.14\,\mathrm{mF}$, $\sigma=1\, \mathrm{S}$, $k=\,4.1667 \times 10^{-5}$.

\emph{Network parameters (Power sharing simulation):} Before the load step: $g_{11}=37.71\,\mathrm{S}$, $g_{22}= 27.87 \,\mathrm{S}$, $g_{33}=50.82 \,\mathrm{S}$, $g_{12}=g_{21}= 8.2 \,\mathrm{S}$, $g_{13}=g_{31}= 24.6 \,\mathrm{S}$, $g_{23}=g_{32}=16.4 \,\mathrm{S}$. After the load step: $g_{11}=37.07 \,\mathrm{S}$, $g_{22}= 27.59 \,\mathrm{S}$, $g_{33}=48.28 \,\mathrm{S}$, $g_{12}=g_{21}= 8.62 \,\mathrm{S}$, $g_{13}=g_{31}= 25.86\,\mathrm{S}$, $g_{23}=g_{32}=17.24 \,\mathrm{S}$.
Parameters correspond to the Kron-reduced network when the load is stepped from $20\Omega$ to $10\Omega$.
\subsection{Details on Averaging Operations}
\label{sec:Avg}
We show in this appendix that by means of averaging and integration by parts, we can derive~\eqref{eq:avgpolar} from~\eqref{eq:averaging3}. We begin by considering the amplitude dynamics. Recall that by standard averaging arguments~\cite[Theorem 10.4]{HKK:02}, we know that the solution of the averaged VOC dynamics~\eqref{eq:averaging1} is $\mathcal{O}(\varepsilon)$ close the solution of original VOC dynamics~\eqref{eq:polar1}  for $s \in [0,t^{*}]$, i.e., $r({\tau}) - \overline r(\varepsilon {\tau}) = \mathcal{O}(\varepsilon)\,.$
From~{\eqref{eq:averaging3}}, we have
\begin{align}
\dot{\overline{r}}(\varepsilon \tau) &- \frac{\varepsilon \alpha}{2}\left(\overline{r}(\varepsilon \tau) - \beta \frac{1}{4} \overline{r}^3(\varepsilon \tau)\right) \nonumber \\ &= -\frac{\varepsilon}{2\pi}\int_{\tau-2\pi}^{\tau}  i(s)\cos(s + \overline \theta)ds \, \nonumber \\
  &=- \frac{\varepsilon}{2\pi}\int_{\tau-2\pi}^{\tau}\frac{1}{\overline r(\varepsilon s)} \overline r(\varepsilon s) i(s)\cos(s + \overline \theta)ds \, \nonumber \\
  &=- \frac{\varepsilon}{2\pi}\int_{\tau-2\pi}^{\tau}\frac{1}{\overline r(\varepsilon s)} \left(r(s)- \mathcal{O}(\varepsilon)\right) i(s) \cos(s + \overline \theta)ds \, \nonumber \\
  &=- \frac{\varepsilon }{2\pi}\int_{\tau-2\pi}^{\tau}\frac{1}{\overline r(\varepsilon s)}r(s)i(s) \cos(s + \overline \theta)ds +\mathcal{O}(\varepsilon^2)\, \nonumber \\
  &=- \frac{\varepsilon}{2\pi}\bigg{[}\frac{1}{\overline r(\varepsilon s)}\int r(s)i(s) \cos(s + \overline \theta)ds \bigg{]}^{\tau}_{\tau-2\pi}\nonumber \\ & - \frac{\varepsilon}{2\pi}\int_{\tau-2\pi}^{\tau}\frac{-\dot {\overline r}(\varepsilon s)}{\overline r^2(\varepsilon s)}\left(\int r(s)i(s) \cos(s + \overline \theta)ds\right)ds \nonumber \\&+\mathcal{O}(\varepsilon^2)\,,
\label{eq:dynamics}  
\end{align}
where the expansion in the last equality leverages integration by parts, { and we changed the integration limits from $[0,2\pi]$ as in \eqref{eq:averaging3} to $[\tau,\tau-2\pi]$ without loss of generality as the integrand is 2$\pi$-periodic.} As $\dot{\overline r}{(\varepsilon s)} = \mathcal{O}(\varepsilon)$, see \eqref{eq:averaging3}, we have
\begin{equation*}
\frac{\varepsilon}{2\pi}\int_{\tau-2\pi}^{\tau}\frac{-\dot {\overline r}{(\varepsilon s)}}{\overline r^2{(\varepsilon s)}}\left(\int r(s)i(s) \cos(s+ \overline \theta)ds\right)ds = \mathcal{O}(\varepsilon^2)\,,
\end{equation*}
and therefore ignoring $\mathcal{O}(\varepsilon^2)$ in~\eqref{eq:dynamics} yields:
\begin{align*}
\dot{\overline{r}}(\varepsilon \tau) &- \frac{\varepsilon \alpha}{2}\left(\overline{r}(\varepsilon \tau) - \beta \frac{1}{4} \overline{r}^3(\varepsilon \tau)\right) \nonumber \\ 
&=- \frac{\varepsilon}{2\pi}\bigg{[}\frac{1}{\overline r(\varepsilon s)}\int r(s)i(s) \cos(s + \overline \theta)ds \bigg{]}^{\tau}_{\tau-2\pi}\,.
\end{align*}
The right-hand side can be further simplified as:
\begin{align}
&\bigg{[}\frac{1}{\overline r(\varepsilon s)}\int r(s)i(s) \cos(s + \overline \theta)ds \bigg{]}^{\tau}_{\tau-2\pi} \nonumber \\ 
&=\frac{1}{\overline r(\varepsilon \tau)} \int r(s)i(s) \cos(s + \overline \theta)ds \bigg{\vert}_{\tau}\nonumber \\ &\quad \quad- \frac{1}{\overline r(\varepsilon (\tau-2\pi))} \int r(s)i(s) \cos(s + \overline \theta)ds \bigg{\vert}_{\tau-2\pi} \, \nonumber \\
&=\frac{1}{\overline r(\varepsilon \tau)} \int r(s)i(s) \cos(s + \overline \theta)ds \bigg{\vert}_{\tau}\nonumber \\ &\quad \quad- \frac{1}{\overline r(\varepsilon \tau) +\mathcal{O}(\varepsilon)} \int r(s)i(s) \cos(s + \overline \theta)ds \bigg{\vert}_{\tau-2\pi} \,, 
\end{align}
where we have the used relation
\begin{equation}
\overline r(\varepsilon \tau)= \overline r (\varepsilon (\tau-2\pi)) + \int_{\tau-2\pi}^{\tau}\dot{\overline r}{(\varepsilon s)}ds = \overline r (\varepsilon (\tau-2\pi)) + \mathcal{O}(\varepsilon)\,, 
\label{eq:first_order}                  
\end{equation}
as $\dot{\overline r}{(\varepsilon s)}= \mathcal{O}(\varepsilon)$. Thus, we can write:
\begin{align}
&\bigg{[}\frac{1}{\overline r(\varepsilon s)}\int r(s)i(s) \cos(s + \overline \theta)ds \bigg{]}^{\tau}_{\tau-2\pi} \nonumber \\ 
&=\frac{1}{\overline r(\varepsilon \tau)}\int r(s)i(s) \cos(s + \overline \theta)ds \bigg{\vert}_{\tau}\nonumber \\&\quad -\frac{1}{\overline r({\varepsilon}\tau)}(1{+}\mathcal{O}(\varepsilon))^{-1}\int r(s)i(s) \cos(s + \overline \theta)ds \bigg{\vert}_{\tau-2\pi}\, \nonumber \\
&=\frac{1}{\overline r(\varepsilon \tau)}\bigg{(}\int r(s)i(s) \cos(s + \overline \theta)ds \bigg{\vert}_{\tau}\nonumber \\ &\quad \quad-\int r(s)i(s) \cos(s + \overline \theta)ds \bigg{\vert}_{\tau-2\pi}\bigg{)} + \mathcal{O}(\varepsilon)\, \nonumber \\
&=\frac{1}{\overline r(\varepsilon \tau)}\int_{\tau-2\pi}^{\tau}r(s)i(s)\cos(s + \overline \theta)ds +\mathcal{O}(\varepsilon)\,\nonumber \\
&=\frac{1}{\overline r(\varepsilon s)}\int_{0}^{2\pi}r(s)i(s) \cos(s + \overline \theta)ds +\mathcal{O}(\varepsilon)\,.
\label{eq:special_term}
\end{align}where we have used the fact that the integrand is 2$\pi$-periodic to change the limits of the definite integral. By substituting~\eqref{eq:special_term} in~\eqref{eq:dynamics} and ignoring $\mathcal{O}(\varepsilon^2)$, we finally arrive at
\begin{equation}
\dot{\overline{r}} = \frac{\varepsilon \alpha}{2}\left(\overline{r}- \beta \frac{1}{4} \overline{r}^3\right)- \frac{\varepsilon}{2\pi \overline r}\int_0^{2\pi}r(\tau)i(\tau) \cos(\tau + \overline \theta)d\tau \,.                  
\label{eq:final_dynamics}
\end{equation}
where we have changed the dummy variable $s$ to $\tau$. Furthermore, by acknowledging that $\theta(\varepsilon )- \overline \theta(\varepsilon \tau)= \mathcal{O}(\varepsilon)$ for $\tau$ in $[0,t^{*}]$, we can expand the cosine function in~\eqref{eq:final_dynamics} as:
\begin{align*}
&\cos(\tau+\overline \theta)=\cos (\tau+\theta-\mathcal{O}(\varepsilon)) \, \nonumber \\
                               &=\cos (\tau+\theta)\cos(\mathcal{O}(\varepsilon))+\sin(\tau+\theta)\sin(\mathcal{O}(\varepsilon))\,\nonumber \\
                               &= \cos (\tau+\theta)(1-\mathcal{O}(\varepsilon^2))+\sin(\tau+\theta)\mathcal{O}(\varepsilon) \,.
\end{align*}
Ignoring $\mathcal{O}(\varepsilon^2)$ terms and transitioning from $\tau$ to $t$ coordinates, we can express~\eqref{eq:final_dynamics} as a function of the average real power as below:
\begin{align}
\frac{d\overline{r}}{dt} &= \frac{\alpha}{2 C} \left(\overline{r} - \beta \frac{1}{4} \overline{r}^3\right) - \frac{\omega}{2\pi C \overline r}\int_0^{\frac{2\pi}{\omega}}  i(t) r(t) \cos(\omega t + {\theta}) dt ,\nonumber \\
&=\frac{\alpha}{2 C} \left(\overline{r} - \beta \frac{1}{4} \overline{r}^3\right) -  \frac{ \omega}{2\pi C \overline r} \overline P \,.             
\end{align}
The phase dynamics can be analysed in a similar manner as follows. From~\eqref{eq:averaging1}, we have
\begin{align*}
\dot{\overline{\theta}}(\varepsilon \tau)&=\frac{\varepsilon}{2\pi}\int_{\tau-2\pi}^{\tau} \frac{1}{\overline{r}(\varepsilon s)}i(s)\sin(s + \overline \theta) ds\, \nonumber \\
&=\frac{\varepsilon}{2\pi}\int_{\tau-2\pi}^{\tau} \frac{\overline r (\varepsilon s)}{\overline{r}^2(\varepsilon s)}i(s)\sin(s + \overline \theta)ds\, \nonumber \\
&=\frac{\varepsilon}{2\pi}\int_{\tau-2\pi}^{\tau}\frac{1}{\overline{r}^2(\varepsilon s)}(r(s)-\mathcal{O}(\varepsilon))i(s)\sin(s + \overline \theta)ds\, \nonumber \\
&=\frac{\varepsilon}{2\pi}\int_{\tau-2\pi}^{\tau}\frac{1}{\overline{r}^2(\varepsilon s)}r(s)i(s)\sin(s + \overline \theta)ds +\mathcal{O}(\varepsilon^2)\, \nonumber \\
&=\frac{\varepsilon}{2\pi}\bigg{[}\frac{1}{\overline{r}^2(\varepsilon \tau)}\int r(s)i(s)\sin(s + \overline \theta)ds\bigg{]}_{\tau-2\pi}^{\tau}\, \nonumber\\&\, -\frac{\varepsilon }{\pi}\int_{\tau-2\pi}^{\tau}\frac{\dot{\overline r}(\varepsilon s)}{\overline r^3(\varepsilon s)}\left(\int r(s)i(s)\sin(s + \overline \theta)d\tau\right)d\tau +\mathcal{O}(\varepsilon^2),
\end{align*}
where the last line again follows from integration by parts. Utilizing the fact that $\dot{\overline r}(\varepsilon \tau) = \mathcal{O}(\varepsilon)$, see \eqref{eq:averaging1}, we have:
\begin{align}
\dot{\overline{\theta}} (\varepsilon \tau)&=\frac{\varepsilon }{2\pi }\bigg{[}\frac{1}{\overline{r}^2(\varepsilon s)}\int r(s)i(s)\sin(s + \overline \theta)d\tau\bigg{]}_{\tau-2\pi}^{\tau} \,, \nonumber \\
&= \frac{\varepsilon }{2\pi}\bigg{(}\frac{1}{\overline{r}^2(\varepsilon \tau)}\int r(s)i(s)\sin(s + \overline \theta)d\tau \bigg{\vert}_{\tau} \nonumber \\
&-\frac{1}{\overline{r}^2(\varepsilon (\tau-2\pi))}\int r(s)i(s)\sin(s + \overline \theta)d\tau \bigg{\vert}_{\tau-2\pi} \bigg{)},
\label{eq: 56}
\end{align}
where we have ignored $\mathcal{O}(\varepsilon^2)$ and higher-order terms. Equation \eqref{eq: 56} can be reformulated by using~\eqref{eq:first_order} as
\begin{align}
\dot{\overline{\theta}}(\varepsilon \tau)
&=\frac{\varepsilon }{2\pi }\bigg{(}\frac{1}{\overline{r}^2(\varepsilon \tau)}\int r(s)i(s)\sin(s + \overline \theta)ds \bigg{\vert}_{\tau} \nonumber \\
&-\frac{(1+\mathcal{O}(\varepsilon))^{-2}}{\overline{r}^2(\varepsilon \tau)}\int r(s)i(s)\sin(s + \overline \theta)ds \bigg{\vert}_{\tau-2\pi} \bigg{)}\,, \nonumber \\
&= \frac{\varepsilon}{2\pi }\frac{1}{\overline r^2(\varepsilon s)}\int_{0}^{2\pi}r(\tau)i\sin(\tau + \overline \theta)d\tau  + \mathcal{O}(\varepsilon^{2})\,.
\label{eq:theta_dynamics}
\end{align}
Again, we can expand sine function in~\eqref{eq:theta_dynamics} as:
\begin{align*}
&\sin(\tau+\overline \theta)=\sin(\tau+\theta-\mathcal{O}(\varepsilon)) \,\nonumber \\
                             &=\sin(\tau+\theta)\cos(\mathcal{O}(\varepsilon))-\cos(\tau+\theta)\sin(\mathcal{O}(\varepsilon))\,\nonumber \\
                             &= \sin(\tau+\theta)(1-\mathcal{O}(\varepsilon^2))-\cos(\tau+\theta)\mathcal{O}(\varepsilon)\,.
\end{align*}
Ignoring $\mathcal{O}(\varepsilon^2)$ terms, we obtain:
\begin{align}
\dot{\overline{\theta}}&=\frac{\varepsilon }{2\pi}\frac{1}{\overline r^2}\int_{0}^{2\pi}r(\tau)i(\tau)\sin(\tau + \theta)d\tau \,.
\end{align}
Transitioning from $\tau$ to $t$ co-ordinates, we thus have:
\begin{align}
\frac{d\overline{\theta}}{dt}&=\frac{\omega}{2\pi C}\frac{1}{\overline r^2}\int_{0}^{2\pi/\omega}r(t)i(t)\sin(\omega t + \theta)dt \nonumber \\
&=\frac{\omega}{2\pi C}\frac{1}{\overline r^2} \overline Q\,.
\end{align}
Thus, we have arrived at \eqref{eq:avgpolar}.
\end{appendix}
\bibliographystyle{ieeetr}
\addcontentsline{toc}{section}{\refname}\bibliography{references}

\begin{thebibliography}{10}

\bibitem{Chandorkar-1993}
M.~C. Chandorkar, D.~M. Divan, and R.~Adapa, ``{Control of parallel connected
  inverters in standalone AC supply systems},'' {\em IEEE Trans. Ind. Appl.},
  vol.~29, pp.~136--143, January 1993.

\bibitem{Zhong13}
Q.-C. Zhong, ``Robust droop controller for accurate proportional load sharing
  among inverters operated in parallel,'' {\em IEEE Trans. Ind. Electron.},
  vol.~60, no.~4, pp.~1281--1290, 2013.

\bibitem{Pogaku_2007}
N.~Pogaku, M.~Prodanovic, and T.~Green, ``Modeling, analysis and testing of
  autonomous operation of an inverter-based microgrid,'' {\em IEEE Trans. Power
  Electron.}, vol.~22, pp.~613--625, March 2007.

\bibitem{Bidram_2012}
A.~Bidram and A.~Davoudi, ``Hierarchical structure of microgrids control
  system,'' {\em IEEE Trans. Smart Grid}, vol.~3, no.~4, pp.~1963--1976, 2012.

\bibitem{SD-SynchTCAS1-2014}
B.~B. Johnson, S.~V. Dhople, A.~O. Hamadeh, and P.~T. Krein, ``{Synchronization
  of Nonlinear Oscillators in an LTI Electrical Power Network},'' {\em IEEE
  Trans. Circuits Syst. I, Reg. Papers}, vol.~61, pp.~834--844, March 2014.

\bibitem{SD-SynchJPV-2014}
B.~B. Johnson, S.~V. Dhople, J.~L. Cale, A.~O. Hamadeh, and P.~T. Krein,
  ``Oscillator-based inverter control for islanded three-phase microgrids,''
  {\em IEEE J. Photovolt.}, vol.~4, pp.~387--395, January 2014.

\bibitem{Dhople-Allerton-2013}
S.~V. Dhople, B.~B. Johnson, and A.~O. Hamadeh, ``{Virtual Oscillator Control
  for voltage source inverters},'' in {\em Allerton Conf. on Communication,
  Control, and Computing}, pp.~1359--1363, October 2013.

\bibitem{SD-SynchTPELS-2014}
B.~B. Johnson, S.~V. Dhople, A.~O. Hamadeh, and P.~T. Krein, ``{Synchronization
  of Parallel Single-Phase Inverters With Virtual Oscillator Control},'' {\em
  IEEE Trans. Power Electron.}, vol.~29, pp.~6124--6138, November 2014.

\bibitem{TorresHespanhaMoehlisSep13}
L.~A.~B. T\^{o}rres, J.~P. Hespanha, and J.~Moehlis, ``Synchronization of
  oscillators coupled through a network with dynamics: A constructive approach
  with applications to the parallel operation of voltage power supplies.''
  under review, Sep. 2013.

\bibitem{TorresHespanhaMoehlisJul12}
L.~A.~B. T\^{o}rres, J.~P. Hespanha, and J.~Moehlis, ``Power supplies dynamical
  synchronization without communication,'' in {\em Proc. of the Power \& Energy
  Society 2012 General Meeting}, July 2012.

\bibitem{RHR-PJH:80}
R.~Rand and P.~Holmes, ``Bifurcation of periodic motions in two weakly coupled
  van der pol oscillators,'' {\em International Journal of Non-Linear
  Mechanics}, vol.~15, no.~4, pp.~387--399, 1980.

\bibitem{Strogatz_Book01}
S.~H. Strogatz, {\em {Nonlinear Dynamics and Chaos: {With} Applications to
  Physics, Biology, Chemistry, and Engineering}}.
\newblock Studies in nonlinearity, Westview Press, 1~ed., Jan. 2001.

\bibitem{HKK:02}
H.~K. Khalil, {\em Nonlinear Systems}.
\newblock Prentice Hall, 3~ed., 2002.

\bibitem{FD-JWSP-FB:14a}
F.~D{\"o}rfler, J.~W. Simpson-Porco, and F.~Bullo, ``{Breaking the Hierarchy:
  Distributed Control \& Economic Optimality in Microgrids},'' 2014.
\newblock Submitted. {Available at {http://arxiv.org/pdf/1401.1767v1.pdf}}.

\bibitem{kundur1994power}
P.~Kundur, N.~J. Balu, and M.~G. Lauby, {\em Power system stability and
  control}, vol.~7.
\newblock McGraw-hill New York, 1994.

\bibitem{Chandorkar13}
J.~M. Guerrero, M.~Chandorkar, T.~Lee, and P.~C. Loh, ``Advanced control
  architectures for intelligent microgrids---{Part I}: {Decentralized} and
  hierarchical control,'' {\em IEEE Trans. Ind. Electron.}, vol.~60,
  pp.~1254--1262, Apr. 2013.

\bibitem{Zhong_Robust13}
Q.-C. Zhong, ``Robust droop controller for accurate proportional load sharing
  among inverters operated in parallel,'' {\em IEEE Trans. Ind. Electron.},
  vol.~60, pp.~1281--1290, April 2013.

\bibitem{Guerrero_Hierarchy11}
J.~M. Guerrero, J.~C. Vasquez, J.~Matas, L.~G. de~Vicu\~{n}a, and M.~Castilla,
  ``Hierarchical control of droop-controlled {AC} and {DC} microgrids--a
  general approach toward standardization,'' {\em IEEE Trans. Ind. Electron.},
  vol.~58, no.~1, pp.~158--172, 2011.

\bibitem{RM-AG-GL-FZ:09}
R.~Majumder, A.~Ghosh, G.~Ledwich, and F.~Zare, ``Angle droop versus frequency
  droop in a voltage source converter based autonomous microgrid,'' in {\em
  IEEE PES General Meeting}, pp.~1--8, July 2009.

\bibitem{SD-BJ-FD-AH:13}
S.~V. Dhople, B.~B. Johnson, F.~D{\"o}rfler, and A.~O. Hamadeh,
  ``Synchronization of nonlinear circuits in dynamic electrical networks with
  general topologies,'' {\em IEEE Trans. Circuits Syst. I, Reg. Papers},
  vol.~61, pp.~2677--2690, September 2014.

\bibitem{Dorfler-13}
F.~D\"{o}rfler and F.~Bullo, ``Kron reduction of graphs with applications to
  electrical networks,'' {\em IEEE Trans. Circuits Syst. I, Reg. Papers},
  vol.~60, pp.~150--163, Jan. 2013.

\bibitem{Mauroy-2012}
A.~Mauroy, P.~Sacr\'{e}, and R.~J. Sepulchre, ``Kick synchronization versus
  diffusive synchronization,'' in {\em IEEE Conf. on Decision and Control},
  pp.~7171--7183, 2012.

\bibitem{tuna2012synchronization}
S.~E. Tuna, ``Synchronization analysis of coupled lienard-type oscillators by
  averaging,'' {\em Automatica}, vol.~48, no.~8, pp.~1885--1891, 2012.

\bibitem{Krein-1989}
P.~T. Krein, J.~Bentsman, R.~M. Bass, and B.~C. Lesieutre, ``On the use of
  averaging for the analysis of power electronic systems,'' in {\em IEEE Power
  Electronics Specialists Conf.}, pp.~463--467, June 1989.

\bibitem{sanders1991generalized}
S.~R. Sanders, J.~M. Noworolski, X.~Z. Liu, and G.~C. Verghese, ``Generalized
  averaging method for power conversion circuits,'' {\em IEEE Trans. Power
  Electron.}, vol.~6, no.~2, pp.~251--259, 1991.

\bibitem{kimball2008singular}
J.~W. Kimball and P.~T. Krein, ``Singular perturbation theory for dc--dc
  converters and application to pfc converters,'' {\em IEEE Trans. Power
  Electron.}, vol.~23, no.~6, pp.~2970--2981, 2008.

\bibitem{lehman1996switching}
B.~Lehman and R.~M. Bass, ``Switching frequency dependent averaged models for
  pwm dc-dc converters,'' {\em IEEE Trans. Power Electron.}, vol.~11, no.~1,
  pp.~89--98, 1996.

\bibitem{caliskan1999multifrequency}
V.~A. Caliskan, O.~Verghese, and A.~M. Stankovic, ``Multifrequency averaging of
  dc/dc converters,'' {\em IEEE Trans. Power Electron.}, vol.~14, no.~1,
  pp.~124--133, 1999.

\bibitem{Dhople_ACC_2015}
M.~Sinha, F.~D\"{o}rfler, B.~B. Johnson, and S.~V. Dhople, ``Virtual oscillator
  control subsumes droop control,'' in {\em American Control Conference},
  submitted 2015.

\bibitem{Simpson-Porco_Synchronization12}
J.~W. Simpson-Porco, F.~D\"orfler, and F.~Bullo, ``Synchronization and power
  sharing for droop-controlled inverters in islanded microgrids,'' {\em
  Automatica}, vol.~49, no.~9, pp.~2603--2611, 2013.

\bibitem{JWSP-FD-FB:13}
J.~W. Simpson-Porco, F.~D{\"o}rfler, and F.~Bullo, ``Voltage stabilization in
  microgrids via quadratic droop control,'','' in {\em IEEE Conf. on Decision
  and Control}, pp.~7582--7589, 2013.

\bibitem{schiffer2013synchronization}
J.~Schiffer, D.~Goldin, J.~Raisch, and T.~Sezi, ``Synchronization of
  droop-controlled microgrids with distributed rotational and electronic
  generation,'' in {\em IEEE Conf. on Decision and Control}, pp.~2334--2339,
  2013.

\bibitem{Wang-Duarte}
F.~Wang, J.~Duarte, and M.~Hendrix, ``Active and reactive power control schemes
  for distributed generation systems under voltage dips,'' in {\em IEEE Energy
  Conversion Congress and Exposition}, pp.~3564--3571, Sept 2009.

\bibitem{Peng-Lai}
F.~Z. Peng and J.-S. Lai, ``Generalized instantaneous reactive power theory for
  three-phase power systems,'' {\em IEEE Trans. Instrum. Meas.}, vol.~45, Feb
  1996.

\bibitem{Zhong}
Q.-C. Zhong and Y.~Zeng, ``Parallel operation of inverters with different types
  of output impedance,'' in {\em Annual Conf. of the IEEE Ind. Electron. Soc.},
  pp.~1398--1403, Nov 2013.
\newblock to appear.

\bibitem{Iravani_Book10}
A.~Yazdani and R.~Iravani, {\em Voltage-Sourced Converters in Power Systems}.
\newblock Hoboken, NJ: John Wiley \& Sons, Inc., 2010.

\bibitem{schiffer2013conditions}
J.~Schiffer, R.~Ortega, A.~Astolfi, J.~Raisch, and T.~Sezi, ``Conditions for
  stability of droop--controlled inverter--based microgrids,'' {\em Automatica.
  Submitted}, 2013.

\bibitem{ortega2002interconnection}
R.~Ortega, A.~Van Der~Schaft, B.~Maschke, and G.~Escobar, ``Interconnection and
  damping assignment passivity-based control of port-controlled hamiltonian
  systems,'' {\em Automatica}, vol.~38, no.~4, pp.~585--596, 2002.

\bibitem{RAH-CRJ:85}
R.~Horn and C.~Johnson, {\em Matrix analysis}.
\newblock Cambridge University Press, 2012.

\bibitem{Carlson_1963}
D.~Carlson and H.~Schneider, ``Inertia theorems for matrices: The semidefinite
  case,'' {\em Jnl. of Mathematical Analysis and Applications}, vol.~6,
  pp.~430--446, 1963.

\bibitem{FD-MC-FB:12c}
F.~D{{\"o}}rfler, M.~Chertkov, and F.~Bullo, ``Synchronization in complex
  oscillator networks and smart grids,'' {\em Proceedings of the National
  Academy of Sciences}, vol.~110, pp.~2005--2010, February 2013.

\bibitem{Dorfler-13-Synch}
F.~D\"{o}rfler and F.~Bullo, ``Synchronization in complex oscillator networks:
  A survey,'' {\em Automatica}, vol.~50, no.~6, pp.~1539--1564, 2014.

\end{thebibliography}
\end{document}